\documentstyle[prb,aps,multicol,epsf,eqsecnum,array]{revtex} 

\begin{document}

\draft 

\title{Electronic transport through ballistic chaotic cavities: reflection
symmetry, direct processes, and symmetry breaking}
\author{Mois\'es Mart\'\i nez \cite{ifunam}}
\address{Departamento de Ciencias B\'asicas, UAM-Azcapotzalco,
C. P. 02200, M\'exico Distrito Federal, M\'exico}
\author{Pier A. Mello}
\address{Instituto de F\'\i sica, Universidad Nacional Aut\'onoma de M\'exico,
01000 M\'exico Distrito Federal, M\'exico}

\date{\today} 

\maketitle 

\begin{abstract}
We extend previous studies on transport through ballistic chaotic cavities 
with spatial left-right (LR) reflection symmetry to include the presence of 
direct processes. We first analyze fully LR-symmetric systems in the 
presence of direct processes and compare the distribution $w(T)$ of the 
transmission coefficient $T$ with that for an asymmetric cavity with the 
same ``optical" $S$ matrix. We then study the problem of ``external mixing" 
of the symmetry caused by an asymmetric coupling of the cavity to the 
outside. We first consider the case where symmetry breaking arises because 
two symmetrically positioned waveguides are coupled to the cavity by means 
of asymmetric tunnel barriers. Although this system is asymmetric with 
respect to the LR operation, it has a striking memory of the symmetry of the 
cavity it was constructed from. Secondly, we break LR symmetry in the 
absence of direct processes by asymmetrically positioning the two waveguides 
and compare the results with those for the completely asymmetric case.
\end{abstract}


\begin{multicols}{2} 
\narrowtext 

\section{Introduction}

\label{introduction}

The problem of chaotic wave scattering is of great interest in various
branches of physics, like optics, nuclear, mesoscopic and microwave physics.

The study of quantum-mechanical scattering problems whose classical dynamics
is chaotic has been further motivated by recent experiments on quantum
electronic transport in microstructures consisting of a cavity connected to 
leads \cite{huibers et al}. We know that symmetries have very interesting 
effects on the properties of the electric conductance in mesoscopic systems: 
time-reversal and spin-rotational symmetries \cite{alw1991,ampzj1995}, as well 
as spatial-reflection symmetries \cite{gmmb,bm1996} have been studied in the 
literature.

The problem of electronic transport through asymmetric (AS) chaotic cavities
is addressed in detail in Ref. \onlinecite{mb1999} in an independent-electron
approximation. In that reference, the possibility of direct processes due to
the presence of short paths is accounted for by specifying the average, or
optical, $S$ matrix $\left\langle S\right\rangle $, within an 
information-theoretic approach. The statistical distribution for the $S$
matrix is known as Poisson's kernel, in which $\left\langle S\right\rangle $
enters as a parameter. When $\left\langle S\right\rangle =0$, i.e. in the
absence of direct processes, the statistical distribution reduces to the
invariant measure for the appropriate universality class.

Microstructures with reflection symmetry and a chaotic classical dynamics
are studied in Refs. \onlinecite{gmmb} and \onlinecite{bm1996}. The analysis is
performed in the absence of direct processes, so that the statistical
distribution of the $S$ matrix is the invariant measure for the universality
class in question and the relevant spatial symmetry: the latter is a
symmetry of the {\it full} system under consideration, i.e. the cavity plus
the two leads that connect the cavity to the outside.

One purpose of the present paper is to extend the study of Refs. \onlinecite{gmmb}
and \onlinecite{bm1996} to include the presence of direct processes. We consider
two-dimensional systems with spinless particles and concentrate on
left-right (LR) symmetry only, i.e. symmetry under reflection through an
axis perpendicular to the current. We also restrict the analysis to
time-reversal-invariant (TRI) problems. One particular way of inducing direct
reflections is by adding potential barriers between the {\it symmetrically
positioned waveguides} and the cavity. If the two barriers are equal, the
system is fully LR symmetric; if the barriers are different, we have a
LR-symmetric cavity coupled {\it asymmetrically} to the outside: using the 
jargon of nuclear physicists \cite{mello67}, we shall refer to this type of
symmetry breaking as ``external mixing'', with an obvious meaning. An
interesting question, amenable to experimental observation, is that of the
interplay between the symmetry of the cavity and external mixing in the
statistical distribution of the conductance of such a structure: the study of 
that interplay is the second main purpose of this paper. From an experimental 
point of view microwave cavities \cite{microwavecav} and acoustic systems
\cite{acousticsystems} might represent good candidates to study these questions.

That interplay may also be there and have interesting effects when $%
\left\langle S\right\rangle =0$, as in the case of a LR-symmetric cavity
coupled to the outside by two waveguides free of potential barriers but
{\it asymetrically located}. This problem can be addressed from the point of view
of the systems described in the previous paragraph in the following way. One
may think of a LR-symmetric cavity coupled to the outside by four waveguides, 
also placed symmetrically. We can break the symmetry by providing the two 
waveguides on the right-hand side of the cavity, say, with identical barriers. The
desired problem is then approached in the limit of impenetrable barriers.

The paper is organized as follows. In order to make the paper reasonably
self-contained, we summarize in the next section a number of concepts that
we shall be using throughout the paper, like the invariant measure and
Poisson's kernel for $S$ matrices and their application to chaotic
scattering in AS cavities, and the invariant measure for LR-symmetric systems. 
Sec. \ref{LR directprocesses} deals with the problem of fully LR-symmetric 
systems in the presence of direct processes. The distribution of the 
conductance is calculated for the particular case of one open channel in each 
lead and a diagonal optical matrix (implying direct reflections), and 
contrasted with the one obtained for an AS chaotic cavity and the same optical 
matrix $\left\langle S\right\rangle $. Different barriers added to the two 
waveguides of an otherwise fully LR-symmetric system with no direct processes 
give rise to direct reflections and external mixing: the problem is studied in 
Sec. \ref{snolr}. Again, the conductance distribution is computed for the 
one-channel case and contrasted with the one obtained for an AS chaotic cavity 
with the same optical matrix $\left\langle S\right\rangle $. The problem of 
external mixing in a LR-symmetric cavity with asymetrically positioned leads 
and $\left\langle S\right\rangle =0$ is addressed in Sec. 
\ref{asymmetric position}. The conductance distribution is calculated and 
compared with the one arising from the invariant measure in the AS case.
Finally, for the sake of completeness, we include a number of appendices where
some of the results mentioned in the text are derived.

\section{The S matrix and its statistical distribution}

\subsection{The scattering problem in the absence of spatial symmetries}
\label{S AS}

\begin{figure}
\input epsf \epsfxsize=8.5cm \centerline{\epsffile{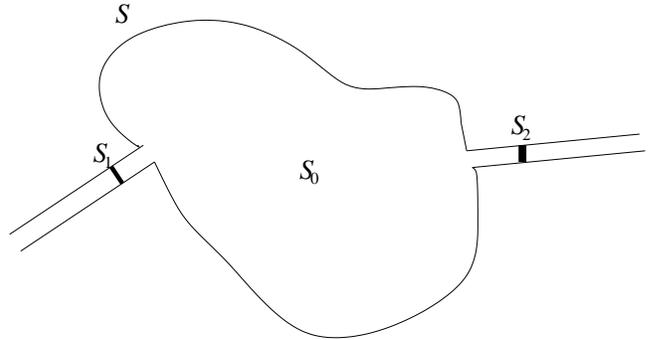}}
\caption{A ballistic chaotic cavity with scattering matrix given by $S_{0}$
connected to two waveguides by means of two barriers with scattering matrices
$S_{1}$ and $S_{2}$.}
\label{cavitybarrier}
\end{figure}

A single-electron scattering problem can be described by the scattering
matrix $S$, which in the stationary case relates the outgoing-wave to the
incoming-wa\-ve amplitudes \cite{n1982}. For a ballistic cavity connected 
to two leads, each with $N$ transverse propagating modes
(see Fig. \ref{cavitybarrier}), the $S$ matrix is $n=2N$-dimensional and 
has the structure 
\begin{equation}
S = \left( 
\begin{array}{cc} 
r & t^{\prime }\\ 
t & r^{\prime }  
\end{array}
\right) \,, \label{sm}
\end{equation}
where $r$, $r^{\prime }$ are the $N\times N$ reflection matrices (for
incidence from either lead) and $t$, $t^{\prime }$ the corresponding
transmission matrices.

From the $S$ matrix we can construct the total transmission coefficient, or 
{\it spinless dimensionless conductance} 
\begin{equation}
T=\mbox{tr}\left( tt^{\dagger }\right) ,  \label{tc}
\end{equation}
which is proportional to the conductance of the cavity, 
\begin{equation}
G=(2e^{2}/h)T,
\end{equation}
the factor 2 arising from the two spin directions.

In Dyson's scheme \cite{dyson1962} there are three basic symmetry classes.
In the absence of any symmetry, the only restriction on $S$ is unitarity,
i.e. 
\begin{equation}
SS^{\dagger }=I,  \label{fc}
\end{equation}
resulting from the physical requirement of flux conservation. This is the
``unitary'' case, also designated as $\beta =2$. For orthogonal symmetry, or 
$\beta =1$, $S$ is symmetric, i.e.
\begin{equation}
S=S^{T},  \label{tri}
\end{equation}
because one has either time-reversal invariance (TRI) and integral spin, or
TRI, half-integral spin and rotational symmetry. In the ``symplectic'' case (%
$\beta =4$), $S$ is self-dual because of TRI with half-integral spin and no
rotational symmetry. From now on we consider the scattering problem of
``spinless'' electrons, so that the case $\beta =4$ will not be touched upon.

A convenient parametrization of the $S$ matrix is the polar representation 
\cite{mpk1988,mp1991}  
\begin{equation}
S=\left( 
\begin{array}{cr}
v_{1} & 0 \\ 
0 & v_{2}
\end{array}
\right) \left( 
\begin{array}{cr}
-\sqrt{1-\tau } & \sqrt{\tau } \\ 
\sqrt{\tau } & \sqrt{1-\tau }
\end{array}
\right) \left( 
\begin{array}{cr}
v_{3} & 0 \\ 
0 & v_{4}
\end{array}
\right) ,  \label{polarrep}
\end{equation} 
where $\tau $ stands for the $N$-dimensional diagonal matrix of eigenvalues 
$\tau _{a}$ ($a=1,\ldots ,N$) of the Hermitian matrix $tt^{\dagger }$; 
$v_{i}$ ($i=1,\ldots ,4$) are arbitrary $N\times N$ unitary matrices for $
\beta =2$, with the restriction $v_{3}=v_{1}^{T}$, $v_{4}=v_{2}^{T}$ for $
\beta =1$.

\subsubsection{The invariant measure}
\label{inv meas}

When the classical dynamics of the system is chaotic, a statistical analysis
of the quantum-mechanical problem is called for. That analysis is performed
in terms of ``ensembles'' of physical systems, described mathematically by
an ensemble of $S$ matrices, endowed with a probability measure. The
starting point to such an analysis is the concept of {\it invariant measure}%
, which is a precise formulation of the intuitive notion of {\it equal a
priori probabilities} in the space of scattering matrices.

The invariant measure, to be designated as $d\mu ^{(\beta )}(S)$, is
invariant under the symmetry operation which is relevant to the universality
class under consideration \cite{dyson1962,hua1963}, i.e. 
\begin{equation}
d\mu ^{(\beta )}(S)=d\mu ^{(\beta )}(U_0 S V_0). \label{inv. meas.}
\end{equation}
Here, $U_{0}$, $V_{0}$ are arbitrary but fixed unitary matrices in the
unitary case, while $V_{0}=U_{0}^{T}$ in the orthogonal one. Eq. (\ref{inv.
meas.}) defines the Circular (Orthogonal, Unitary) Ensembles (COE, CUE), for 
$\beta =1,2$, respectively.

\subsubsection{Chaotic scattering by AS cavities} \label{AScavities}

The information-theoretic approach of Refs. \onlinecite{mps85,fm1985} leads 
to the
probability distribution known as Poisson's kernel \cite{mb1999,hua1963}  
\begin{equation}
dP_{\langle S\rangle }^{(\beta )}(S) = 
\frac{[\det (I-\langle S\rangle \langle 
S\rangle ^{\dagger })]^{(\beta n+2-\beta )/2}}  
{|\det (I-S\langle S\rangle ^{\dagger })|^{\beta n+2-\beta }}  
d\mu ^{(\beta )}(S),  \label{pk1}
\end{equation}
where the invariant measure is assumed normalized, i.e. 
\begin{equation}
\int d\mu ^{(\beta )}(S)=1.  \label{norm inv meas}
\end{equation}
Here, $n=2N$ is the dimensionality of the $S$ matrix and $\langle S\rangle $
is the averaged, or {\it optical,} $S$ matrix, which describes the promt
response arising from {\it direct processes}.

In the absence of direct processes, $\langle S\rangle =0$ and Poisson's
measure (\ref{pk1}) reduces to the invariant measure for the universality  
class in question. In terms of the polar representation, the invariant  
measure can be written as \cite{bm1994,jpb1994} 
\begin{equation}
d\mu ^{(\beta )}\left( S\right) =p^{(\beta )}\left( \{\tau \}\right)
\prod_{a}d\tau _{a}\prod_{i}d\mu \left( v_{i}\right) \,.
\label{invariantmeasure}
\end{equation}
Here, the joint probability density of \{$\tau $\} is 
\begin{equation}
p^{(\beta )}\left( \{\tau \}\right) =C_{\beta }\prod_{a<b}|\tau _{a}-\tau
_{b}|^{\beta }\prod_{c}\tau _{c}^{(\beta -2)/2},  \label{jointprobtau}
\end{equation}
$C_{\beta }$ being a normalization constant and $d\mu (v_{i})$ denoting the
invariant measure on the unitary group $U(N)$ for matrices $v_i$.

For $\langle S\rangle \neq 0$, a useful construction of Poisson's ensemble
is given in Refs. \onlinecite{brower1,brower2}. Consider the system shown in 
Fig. \ref{cavitybarrier}: it consists of a cavity described by the $n$%
-dimensional scattering matrix $S_{0},$ connected to two leads by the tunnel
barriers described by the $n\times n$ scattering matrices
\begin{eqnarray}
S_{1} &=&\left( 
\begin{array}{cr}
r_{1} & t_{1}^{\prime } \\ 
t_{1} & r_{1}^{\prime }
\end{array}
\right) ,  \label{S1} \\
S_{2} &=&\left( 
\begin{array}{cr}
r_{2} & t_{2}^{\prime } \\ 
t_{2} & r_{2}^{\prime }
\end{array}
\right) ,  \label{S2}
\end{eqnarray}
respectively. We bunch the two leads into a ``superlead'' and construct the 
$2n\times 2n$ scattering matrix $S_{b}$ 
\begin{equation}
S_{b}=\left( 
\begin{array}{cr}
r_{b} & t_{b}^{\prime } \\ 
t_{b} & r_{b}^{\prime }
\end{array}
\right) =\left( 
\begin{array}{llll}
r_{1} & 0 & t_{1}^{\prime } & 0 \\ 
0 & r_{2}^{\prime } & 0 & t_{2} \\ 
t_{1} & 0 & r_{1}^{\prime } & 0 \\ 
0 & t_{2}^{\prime } & 0 & r_{2}
\end{array}
\right) \quad .  \label{sbarriers}
\end{equation}
Here, the various blocks ($r_{b}$, etc.) are $n$-dimensional. The scattering
matrix $S_{0}$ for the cavity can be written in terms of the scattering
matrix $S$ for the full system $\left\{ {\rm cavity}+{\rm barriers}\right\} $
as 
\begin{equation}
S_{0}=\frac{1}{t_{b}^{\prime }}\left( S-r_{b}\right) \frac{1}{%
I-r_{b}^{\dagger }S}t_{b}^{\dagger }.  \label{S0(S)}
\end{equation}
One can prove \cite{hua1963,fm1985,brower1,brower2} that between the
invariant measures for $S_{0}$ and for $S$ we have the Jacobian
\begin{equation}
d\mu ^{(\beta )}(S_{0})=\frac{[\det (I-\langle S\rangle \langle S\rangle
^{\dagger })]^{(\beta n+2-\beta )/2}}{|\det (I-S\langle S\rangle ^{\dagger
})|^{\beta n+2-\beta }}d\mu ^{(\beta )}(S) . \label{jacobian(S0,S)}
\end{equation}
Now, if the matrix $S_{0}$ for the cavity is distributed according to
the invariant measure, i.e. $d\mu ^{(\beta )}(S_{0})$, the distribution of
the transformed $S$ satisfies 
\begin{equation}
dP(S)=d\mu ^{(\beta )}(S_{0})  \label{dP(S)dmu(S0)}
\end{equation}
and we obtain Eq. (\ref{pk1}), the optical $S$ being given by
the $n$-dimensional matrix 
\begin{equation}
\langle S\rangle =r_{b}=\left( 
\begin{array}{ll}
r_{1} & 0 \\ 
0 & r_{2}^{\prime }
\end{array}
\right) .  \label{av(S)=rb}
\end{equation}

{\it The $N=1$, $\beta =1$ case. The $T$ distribution.}
We now consider the distribution of the $S$ matrix for the system shown in Fig.
\ref{cavitybarrier} for the case $N=1$ and $\beta =1$. The matrices $S_{0}$
of the ballistic cavity, $S_{1}$ and $S_{2}$ of the two tunnel barriers and  
$S$ [related through Eq. (\ref{S0(S)})] are $2\times 2$ and have the
structure (\ref{sm}) with $t^{\prime }=t$. In the polar representation (\ref
{polarrep}) we have three independent parameters $\tau $, $\phi $, $\psi $,
where we have written $v_{1}=e^{i\phi }$, $v_{2}=e^{i\psi }$. The range of
variation of these parameters is taken to be
\begin{equation}
\begin{array}{rl}
\tau  &\in [0,1], \\
\phi ,\psi  &\in [0,2\pi ].
\end{array} \label{range}
\end{equation}
In terms of them, $S$ can be written as 
\begin{equation}
S=\left( 
\begin{array}{cr}
r & t \\ 
t & r^{\prime }
\end{array}
\right) =\left[ 
\begin{array}{cr}
-\sqrt{1-\tau }\,e^{2i\phi } & \sqrt{\tau }\,e^{i(\phi +\psi )} \\ 
\sqrt{\tau }\,e^{i(\phi +\psi )} & \sqrt{1-\tau }\,e^{2i\psi }
\end{array}
\right] \,.  \label{S polar N1}
\end{equation}
and the invariant measure of Eqs. (\ref{invariantmeasure}) and (\ref
{jointprobtau}) as 
\begin{equation}
d\mu (S)=\,\frac{d\tau }{2\sqrt{\tau }}\,\frac{d\phi }{2\pi }\,\frac{d\psi }
{2\pi }.  \label{inv meas N1}
\end{equation}

The distribution of $S$ is given by Poisson's kernel, with the optical $S$
matrix 
\begin{equation}
\langle S\rangle =r_{b}=\left( 
\begin{array}{cr}
r_{1} & 0 \\ 
0 & r_{2}^{\prime }
\end{array}
\right) \,.  \label{opticals}
\end{equation}
Substituting $\langle S\rangle $ in Eq. (\ref{pk1}), Poisson's measure can 
be written as 
\begin{equation}
dP_{r_{1},r_{2}^{\prime }}(S)=\frac{\left[ \left( 1-|r_{1}|^{2}\right)
\left( 1-|r_{2}^{\prime }|^{2}\right) \right] ^{3/2}}{\left| \left(
1-rr_{1}^{*}\right) \left( 1-r^{\prime }r{_{2}^{\prime }}^{*}\right) -t^{2}{%
r_{1}}^{*}{r_{2}^{\prime }}^{*}\right| ^{3}}\,d\mu (S)\,.  \label{pkexample1}
\end{equation}

By definition, the resulting distribution of the transmission coefficient $T$
can be expressed as the integral 
\begin{equation}
w_{r_{1},r_{2}^{\prime }}(T)=\int \delta \left( T-\tau \right)
dP_{r_{1},r_{2}^{\prime }}(S).  \label{w(T) 0}
\end{equation}
For this distribution, Ref. \onlinecite{mb1999} gives the expression 
\begin{equation}
\begin{array}{ll}
w_{r_1, r'_2}(T) = \frac{1}{2\sqrt{T}} \left[ \left( 1-\left| r_1
\right|^2 \right) \left( 1-\left| r'_2 \right|^2 \right) \right]^{3/2}
&  \\
\times  \left\langle \frac{1}{\left| \left( e^{-i\varphi }+\left|
r_1 \right| \sqrt{1-T}\right) \left( e^{-i\psi }+\left| r'_2
\right| \sqrt{1-T}\right) -\left| r_1 \right| \left| r'_2
\right| T\right|^3 } \right\rangle _{\varphi ,\psi }, &
\end{array} \label{w(T) 1}
\end{equation}
where $\left\langle \cdot \cdot \cdot \right\rangle _{\varphi ,\psi }$
denotes an average over the variables $\varphi $ and $\psi $ over the
interval $\left[ 0,2\pi \right] $. When $r_{1}=r_{2}^{\prime }=0$, the above
expression (\ref{w(T) 1}) reduces to  
\begin{equation}\label{w00(T)}  
w_{0,0}(T) = \frac{1}{2\sqrt{T}} \, , 
\end{equation}  
as it should. Fig. \ref{wtsimrr} ahead (Sect. \ref{LR directprocesses}) 
shows with dotted lines the evolution of $w_{r_1, r'_2}(T)$ for 
$r_1 = r'_2 = \langle r\rangle$ with the parameter $\langle r\rangle$,
obtained from Eq. (\ref{w(T) 1}) by numerical integration. That distribution 
tends to $\delta (T)$ as $\langle r\rangle \longrightarrow 0$.  

To further illustrate the physics resulting from the $S$-matrix distribution 
(\ref{pkexample1}) we analyze the special case $r_{1}=0$, so that the right
barrier is the only one present. For this case, Eqs. (\ref{pkexample1}), 
(\ref{inv meas N1}) give, for the joint probability distribution of the
parameters $\tau $, $\phi $, $\psi $, the expression 
\begin{equation}
dP_{0,r_{2}^{\prime }}(S)=\frac{\left( 1-|r_{2}^{\prime }|^{2}\right) ^{3/2}%
}{\left| 1-\sqrt{1-\tau }\,e^{2i\psi }\,r{_{2}^{\prime }}^{*}\right| ^{3}}\,%
\frac{d\tau }{2\sqrt{\tau }}\,\frac{d\phi }{2\pi }\,\frac{d\psi }{2\pi }.
\label{pkexample2}
\end{equation}
We first notice that the angular variable $\phi $ is {\it uniformly distributed}
for all $r{_{2}^{\prime }}$. In this particular case the $T$ probability
density of Eq. (\ref{w(T) 1}) can be integrated analytically, to give \cite
{mb1999} 
\begin{equation}
w_{0,r_{2}^{\prime }}(T)=\frac{\left( 1-\left| r_{2}^{\prime }\right|
^{2}\right) ^{3/2}}{2\sqrt{T}}\left. _{2}F_{1}\left( 3/2;3/2;1;\left|
r_{2}^{\prime }\right| ^{2}\left( 1-T\right) \right) \right.,
\label{w(T) 2}
\end{equation}
$_{2}F_{1}$ being a hypergeometric function \cite{abram}.

As a check, we consider two limiting situations. Firstly, for $r_{2}^{\prime
}=0$ we have a ballistic cavity without prompt response. The probability
distribution for $S$, $dP_{0,0}(S)$ [see Eq. (\ref{pkexample2})], goes back 
to the invariant measure (\ref{inv meas N1}), as it should. Secondly, we 
obstruct the right lead by making the barrier there a perfect reflector. As 
a result, $r'_2=-1$ and it can be shown 
(see Appendix \ref{ProofASdP(0,-1)(S)}) that 
$dP_{0,r_{2}^{\prime}}(S)$ reduces to 
\begin{equation}
dP_{0,-1}(S)=\delta \left( \tau \right) d\tau \frac{d\phi }{2\pi } 
\frac{1}{2}\left[ \delta \left( \psi -\frac{\pi }{2}\right) + 
\delta \left( \psi - 3\frac{\pi }{2}\right) \right] d\psi ,  
\label{ASdP(0,-1)(S)} 
\end{equation}  
where the angles in the arguments of the delta functions are defined modulo  
$2\pi$. We see from the above expression that the distribution of $\tau $ is 
a one-sided delta function at zero, i.e.
\begin{equation}
w(T)=\delta (T) ,  \label{w(T)}
\end{equation}
so that the transmission tends to zero, as expected.
Also, the distribution of $\psi $ consists of delta functions centered at $%
\pi /2$ and 3$\pi /2$, so as to ensure the vanishing of the wave function at
the impenetrable barrier. In contrast, as already noted, the variable $\phi $
is uniformly distributed from 0 to $2\pi $. In this limiting case we end up
with a ballistic cavity connected to just one lead: thus {\it the resulting
1-dimensional $S$ matrix $r=-e^{2i\phi }$ is distributed according to the
invariant measure}.

Now we go back to the intermediate case in which $r'_2$ in Eq. 
(\ref{w(T) 2}) is real and $-1<r'_2<0$. We show in Fig. \ref{wtsim0rp2} 
ahead (Sect. \ref{snolr}) with dotted lines the evolution of the $T$ 
distribution for several values of $r_{2}^{\prime }$, obtained from the 
analytical result (\ref{w(T) 2}). 

\subsection{The scattering problem for TRI, LR-symmetric systems}
\label{S LR av(S)0}

In the presence of additional symmetries, for fixed values for all quantum
numbers of the full symmetry group the invariant ensemble is one of the
three circular ensembles in Dyson's scheme. Thus for reflection symmetric
systems $S$ is block diagonal in a basis of definite parity with respect to
reflections, with a circular ensemble in each block \cite{gmmb,bm1996}.

For a system with TRI and LR symmetry the general form of the $S$ matrix is
\begin{equation}
S=\left( 
\begin{array}{cr}
r & t \\ 
t & r
\end{array}
\right) ,  \label{lrs}
\end{equation}
with
\begin{mathletters}
\begin{eqnarray}
r & = & r^{T} \\ t & = & t^T .
\end{eqnarray} \label{rtsymm}
\end{mathletters}

All the matrices with the structure (\ref{lrs}) can be simultaneoulsy
brought to block-diagonal form using the rotation matrix 
\begin{equation}
R_{0}=\frac{1}{\sqrt{2}}\left( 
\begin{array}{rr}
I_{N} & I_{N} \\ 
-I_{N} & I_{N}
\end{array}
\right) ,  \label{rotation}
\end{equation}
where $I_{N}$ is the $N$-dimensional unit matrix. In fact 
\begin{equation}
S^{\prime }=R_{0}SR_{0}^{T}=\left[ 
\begin{array}{cr}
s^{(+)} & 0 \\ 
0 & s^{(-)}
\end{array}
\right] ,  \label{Srot}
\end{equation}
with 
\begin{equation}
s^{(\pm )}=r\pm t.  \label{s+-}
\end{equation}
Since $S$ is unitary and symmetric, so are $S^{\prime }$ and the two $%
N\times N$ matrices $s^{(\pm )}$. While $S$ has the restricted form (\ref
{lrs}), $s^{(\pm )}$ are the {\it most general }$N\times N${\it \ unitary
and symmetric}, i.e. $\beta =1$, matrices.

\subsubsection{The invariant measure}
\label{inv meas LR}

The invariant measure for $S$ matrices with the structure (\ref{lrs}) was
found in Refs. \onlinecite{gmmb,bm1996}, based on the consideration that two
arbitrary unitary symmetric matrices $s^{(\pm )}$ can generate the most
general unitary $S$ matrix with the structure (\ref{lrs}). The invariant
measure for matrices of the form (\ref{lrs}) can be written as 
\begin{equation}
d\widehat{\mu }^{(1)}(S)=d\mu ^{(1)}(s^{(+)})d\mu ^{(1)}(s^{(-)}),
\label{invmeasLR}
\end{equation}
where $d\mu ^{(1)}(s^{(\pm )})$ is the invariant measure discussed above for
unitary and symmetric matrices ($\beta =1$) in the absence of spatial
symmetries.

\subsubsection{Chaotic scattering by systems with full LR symmetry in the
absence of direct processes}

It has been found \cite{bm1996} that single-electron scattering by
classically chaotic cavities with LR symmetry and in the absence of direct
processes is well described by the invariant measure discussed above.

{\it The $N=1$ case. The $T$ distribution.}
Ref. \onlinecite{gmmb} finds the distribution of the total transmission
coefficient $T$ for the one-channel case ($N=1$) arising from the invariant 
measure (\ref{invmeasLR}) as 
\begin{equation}
w(T)=\frac{1}{\pi \sqrt{T(1-T)}}.  \label{w(T)LR}
\end{equation}

\section{Systems with TRI and full LR symmetry in the presence of direct
processes}
\label{LR directprocesses}

In this section we study a TRI system with full LR symmetry, just as in Sec. 
\ref{S LR av(S)0}, but now admitting the possibility of direct processes.
For the systems analyzed in Sec. \ref{S LR av(S)0}, the average (or optical) 
$S$ matrix $\left\langle S\right\rangle $ vanishes, indicating the absence
of a prompt response, whereas now $\left\langle S\right\rangle \neq 0$.

The $S$ matrix has the structure of Eq. (\ref{lrs}), and so does
$\left\langle S\right\rangle $, i.e.
\begin{equation}
\left\langle S\right\rangle =\left( 
\begin{array}{cr}
\left\langle r\right\rangle & \left\langle t\right\rangle \\ 
\left\langle t\right\rangle & \left\langle r\right\rangle
\end{array}
\right) \,,  \label{av(lrs)}
\end{equation}
with 
\begin{mathletters}
\begin{eqnarray}
\left\langle r\right\rangle &=&\left\langle r\right\rangle ^{T} \\
\left\langle t\right\rangle &=&\left\langle t\right\rangle ^{T}
\end{eqnarray}\label{av(r,t)symm}
\end{mathletters}
being $N\times N$ blocks. Both $S$ and $\left\langle S\right\rangle $ can be
brought to a block-diagonal form by the rotation matrix (\ref{rotation}): $S$
becomes $S^{\prime }$ of Eq. (\ref{Srot}) and $\left\langle S\right\rangle $
becomes
\begin{equation}
\left\langle S^{\prime }\right\rangle =R_{0}\left\langle S\right\rangle
R_{0}^{T}=\left[ 
\begin{array}{cr}
\left\langle s^{(+)}\right\rangle & 0 \\ 
0 & \left\langle s^{(-)}\right\rangle
\end{array}
\right] .  \label{av(S)rot}
\end{equation}

As we noticed right below Eq. (\ref{s+-}), $s^{(\pm )}$ are the {\it most
general }$N\times N${\it \ unitary and symmetric} matrices; they thus belong
to the $\beta =1$ universality class. Their distribution is given by two
statistically independent Poisson's kernels of the form (\ref{pk1}), 
with $\left\langle s^{(\pm )}\right\rangle $ as their optical
matrices. Denoting by $d\widehat{P}_{\langle S\rangle}(S)$ the $S$ matrix
distribution, we have
\begin{equation}
d\widehat{P}_{\langle S\rangle }(S) = dP_{\left\langle s^{(+)}\right\rangle
}(s^{(+)})\,dP_{\left\langle s^{(-)}\right\rangle }(s^{(-)})\,,
\label{lrkernel1}
\end{equation}
where 
\begin{eqnarray}  
&&dP_{\left\langle s^{(\pm)}\right\rangle }(s^{(\pm )}) \nonumber \\ 
&& = \frac{\left[ \det \left( I_{N}-\left\langle s^{(\pm )}\right\rangle
\left\langle s^{(\pm )}\right\rangle ^{\dagger }\right) \right]
^{(N+1)/2}}{\left| \det \left( I_{N}-s^{(\pm )}\left\langle s^{(\pm
)}\right\rangle ^{\dagger }\right) \right| ^{N+1}}\,d\mu ^{(1)}(s^{(\pm
)}) .  \label{pkspm} 
\end{eqnarray} 
We can thus write 
$d\widehat{P}_{\langle S\rangle }(S)$ as  
\begin{eqnarray}
&&d\widehat{P}_{\langle S\rangle }(S) = \frac{ \left[ \det \left(
I_{N}-\left\langle s^{(+)}\right\rangle \left\langle s^{(+)} 
\right\rangle^{\dagger }\right) \right] ^{(N+1)/2}} 
{\left| \det \left( I_{N}-s^{(+)} \left\langle s^{(+)} 
\right\rangle ^{\dagger} \right) \right| ^{N+1}} \nonumber \\ 
&& \times \frac{\left[ \det \left( I_{N}-\left\langle s^{(-)}\right\rangle
\left\langle s^{(-)}\right\rangle ^{\dagger }\right) \right] ^{(N+1)/2}} 
{\left| \det \left( I_{N}-s^{(-)}\left\langle s^{(-)}\right\rangle ^{\dagger
}\right) \right| ^{N+1}}\,d\hat{\mu}^{(1)}(S)  \label{lrkernel2}
\end{eqnarray}
where $d\hat{\mu}^{(1)}(S)$ is defined in Eq. (\ref{invmeasLR}).

The special case of no direct transmission,  
$\left\langle t\right\rangle =0$, i.e.  
\begin{equation}
\left\langle S\right\rangle =\left( 
\begin{array}{cr}
\left\langle r\right\rangle  & 0 \\ 
0 & \left\langle r\right\rangle 
\end{array}
\right) ,  \label{av(S)t0}
\end{equation}
can be written as 
\begin{eqnarray}
&& d\widehat{P}_{\langle r\rangle }(S) \nonumber \\ 
&& = \frac{\left[ \det \left( I_{N}-\left\langle r\right\rangle 
\left\langle r\right\rangle ^{\dagger }\right) \right] ^{N+1}}{\left| \det
\left( I_{N}-s^{(+)}\left\langle r\right\rangle ^{\dagger }\right) 
\right|^{N+1}\left| \det \left( I_{N}-s^{(-)}\left\langle r 
\right\rangle^{\dagger }\right) \right| ^{N+1}} \nonumber \\ 
&& \times \,\, d\hat{\mu}^{(1)}(S) \, .  \label{PkernLRt0}  
\end{eqnarray}
Physically, this case could be realized by fully LR-symmetric structures with
no direct processes, to which identical barriers (with the $\beta =1$
symmetry) are added in the two leads, each with a reflection matrix
[see Eq. (\ref{av(S)=rb})]
\begin{equation}
r_1 = r'_2 = \left\langle r \right\rangle .
\label{rbarr}
\end{equation}
The situation is illustrated in Fig. \ref{cavitylr} ahead but with equals
barriers.

\subsection{The $N=1$ case. The $T$ distribution.}
\label{w(T) N1 LRdir}

In this case, Eq. (\ref{PkernLRt0}) reduces to
\begin{equation}
d\widehat{P}_{\langle r\rangle }(S)=\frac{\left[ 1-\left\langle
r\right\rangle \left\langle r\right\rangle ^{*}\right] ^{2}}{\left|
1-s^{(+)}\left\langle r\right\rangle ^{*}\right| ^{2}\left|
1-s^{(-)}\left\langle r\right\rangle ^{*}\right| ^{2}}\,\,\,d\hat{\mu}%
^{(1)}(S),  \label{PkernLRt0N1}
\end{equation}
where $\left\langle r\right\rangle $ and $s^{(\pm )}$ are now $1\times 1$
matrices, i.e. just complex numbers. The distribution of $T$ can be obtained
from the general expression (\ref{w(T) 0}). For $\langle r\rangle$  
{\it real}, some of the relevant steps are found in App.  
\ref{DerivEqw(T)LRav(r) 1}, the final result being 
\begin{equation}
w_{\langle r\rangle}(T)=\frac{1}{\pi \sqrt{T\left( 1-T\right) }} 
\frac{\left( 1+{\langle r\rangle}^{2}\right) 
\left( 1-{\langle r\rangle}^{2}\right) } 
{\left( 1+{\langle r\rangle}^{2}\right)^{2}-4{\langle r\rangle}^{2} 
\left( 1-T\right) }.  \label{w(T)LRav(r) 1} 
\end{equation}

\begin{figure}[tbh]
\input epsf \epsfxsize=8.5cm \centerline{\epsffile{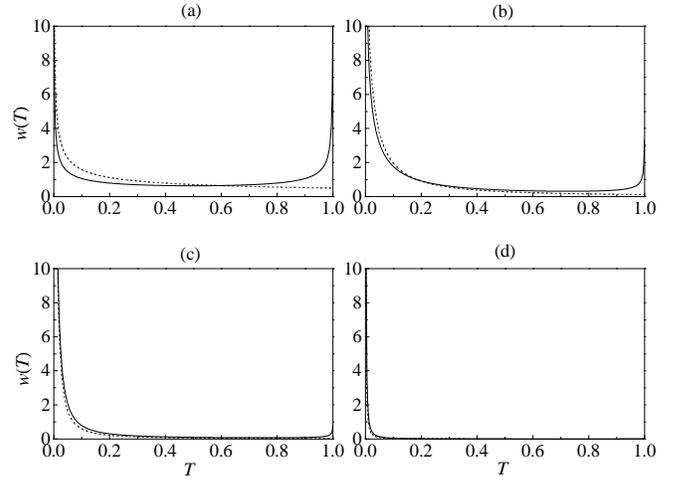}}  
\caption{Shown with a heavy line is the evolution of the distribution  
$w_{\langle r\rangle}(T)$ of Eq. (\ref{w(T)LRav(r) 1}) with the parameter  
$\langle r\rangle = -\cos \epsilon$ for a chaotic cavity with full LR  
symmetry. The cases $\epsilon = \frac{\pi}{2}, \frac{\pi}{4}, \frac{\pi}{8},  
\frac{\pi}{32}$ are shown in (a), (b), (c), (d), respectively. The dotted  
lines show for comparison the $T$ distribution corresponding to an AS cavity
with two identical barriers.} \label{wtsimrr}  
\end{figure}

The distribution (\ref{w(T)LRav(r) 1}) is plotted in Fig. \ref{wtsimrr}  
for several values of $\left\langle r\right\rangle $ and compared, in the
same figure, with the distribution corresponding to an AS cavity with the  
same $\left\langle S\right\rangle $, as given by Eq. (\ref{w(T) 1}).

For $\langle r\rangle=0$, the distribution of Eq. (\ref{w(T)LRav(r) 1})  
reduces to that of Eq. (\ref{w(T)LR}), which is symmetric with respect to 
$T=\frac{1}{2}$, so that $T$ and $R=1-T$ are identically distributed; this
feature is lost when $\langle r\rangle \neq 0$, as small $T$'s become more 
probable. As $\langle r\rangle \longrightarrow -1$, both distributions shown
in the figure (i.e. for LR-symmetric and AS systems) tend to $\delta(T)$.

\section{Breaking the reflection symmetry of cavities by direct processes}
\label{snolr}

\begin{figure}[tbp]
\input epsf \epsfxsize=8.5cm \centerline{\epsffile{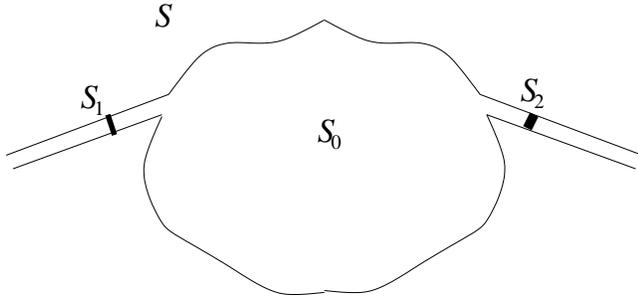}}
\caption{A ballistic cavity with reflection symmetry described by the matrix 
$S_0$, connected to two waveguides by means of two barriers described by
$S_{1}$, $S_{2}$. The barriers give rise to direct processes and, if they are
different, the LR symmetry of the full system is broken (external mixing).}
\label{cavitylr}  
\end{figure}

In this section we study a TRI configuration consisting of a ballistic cavity
with LR symmetry and scattering matrix $S_0$, connected to two symmetrically
positioned waveguides by means of barriers described by $S_1$ and $S_2$,
respectively; in general, the barriers are allowed to be different. This
arrangement introduces direct reflections and a {\it breakdown of the reflection
symmetry} (see Fig. \ref{cavitylr}). As a result, while the scattering matrix
$S_{0}$ of the cavity plus the symmetrically positioned waveguides, but not
including the barriers, has the restricted structure (\ref{lrs}), (\ref{rtsymm}),
the scattering matrix $S$ of the total system including the barriers has the more
general form (\ref{sm}), (\ref{tri}). Now, $S$ is generated from $S_{0}$ through
the inverse of the relation (\ref{S0(S)}); thus, varying $S_{0}$ across its
manifold of independent parameters, but keeping the barriers fixed, generates a
matrix $S$ that varies over a manifold with the same dimensionality. In what
follows we restrict ourselves to the one-channel case ($N=1$) in each lead. The
matrices $S_{0}$ can be expressed in terms of {\it two} independent continuous
parameters (plus a discrete parameter $\sigma $), as in Eq. (\ref{S0 polar 1})
below, while $S$ has the more general form (\ref{S polar N1}); thus there
should be an algebraic relation connecting the three continuous parameters
$\tau $, $\phi $, $\psi $ appearing in the latter equation.

We want $S_{0}$ to be distributed according to the invariant measure
$d\widehat{\mu }(S_{0})$. In principle, the transformation between $S_{0}$ and
$S$ (for fixed $S_{1}$ and $S_2$) defines uniquely the resulting statistical
distribution of $S$, to be called $d\widehat P(S)$
[see Eq. (\ref{dhatP(S)dmu(S0)})]; for that purpose one could find the Jacobian of
the transformation relating $S$ to $S_0$, both matrices being subject to the
restrictions explained in the previous paragraph. In what follows, though, we find
it convenient to compute $d\widehat P(S)$ proceeding along a simpler route, taking
advantage of the Jacobian between {\it unrestricted} $S$ matrices that we already
know from Eq. (\ref{jacobian(S0,S)}). In fact, the measure
$d\widehat{\mu } (S_{0})$ can be first expressed as the measure $d\mu (S_{0})$  of
{\it unrestricted} $S_{0}$ matrices of the form of Eq. (\ref{S0 polar}) below,
times the appropriate delta functions that provide the required restriction
[see Eq. (\ref{psi(phi) 1})] among the three parameters $\tau _{0}$, $\phi _{0}$,
$\psi _{0}$. Next, Eq. (\ref{jacobian(S0,S)}) expresses $d\mu (S_{0})$ in terms of
$d\mu (S)$, the factor in front of $d\mu (S)$ in Eq. (\ref{jacobian(S0,S)}) being
the Jacobian of the transformation from unrestricted $S_{0}$ to unrestricted $S$
matrices. Finally, the identity (\ref{dhatP(S)dmu(S0)}) gives the required
distribution $d\widehat P(S)$ for the $S$ matrices. We proceed to implement this
scheme in detail.
 
The relation between the scattering matrix $S_{0}$ for the cavity and the
matrix $S$ for the full system is given by Eq. (\ref{S0(S)}), with 
\begin{equation}
t_{b}=t_{b}^{\prime }=\left( 
\begin{array}{cr}
t_{1} & 0 \\ 
0 & t_{2}
\end{array}
\right) \,,  \label{tb}
\end{equation}
\begin{equation}
r_{b}=\left( 
\begin{array}{cr}
r_{1} & 0 \\ 
0 & r_{2}^{\prime }
\end{array}
\right) ,\qquad r_{b}^{\prime }=\left( 
\begin{array}{cr}
r_{1}^{\prime } & 0 \\ 
0 & r_{2}
\end{array}
\right) .  \label{rb}
\end{equation}
Here all the matrices are 2-dimensional, so that the various entries are
just complex numbers. We thus have
\begin{equation}
S_{0}={t_{b}}^{-1}\,(S-r_{b})\,\frac{1}{I_{2}-{r_{b}}^{\dagger }\,S}\,{t_{b}}%
^{\dagger }.  \label{lrs02}
\end{equation}
The matrix $S$ has the structure (\ref{S polar N1}), while $S_{0}$ has the
structure (\ref{lrs}), i.e.  
\begin{equation}
S_{0}=\left( 
\begin{array}{cr}
r_{0} & t_{0} \\ 
t_{0} & r_{0}
\end{array}
\right) .  \label{S0 N1}
\end{equation}

It will be useful to write $S_{0}$ of Eq. (\ref{S0 N1}) in the polar
representation (\ref{S polar N1}) as 
\begin{equation}
S_{0}=\left[ 
\begin{array}{cr}
-\sqrt{1-\tau _{0}}\,e^{2i\phi _{0}} & \sqrt{\tau _{0}}\,e^{i(\phi _{0}+\psi
_{0})} \\ 
\sqrt{\tau _{0}}\,e^{i(\phi _{0}+\psi _{0})} & \sqrt{1-\tau _{0}}\,e^{2i\psi
_{0}}
\end{array}
\right] .  \label{S0 polar}
\end{equation}
However, the three parameters $\tau _{0}$, $\phi _{0}$ and $\psi _{0}$ are
not independent. In fact, the structure of $S_{0}$ given in Eq. (\ref{S0 N1}%
) implies a relation between the two angles $\phi _{0}$ and $\psi _{0}$,
i.e. 
\begin{equation}
e^{2i\psi _{0}}=-e^{2i\phi _{0}},  \label{psi(phi) 1}
\end{equation}
or, taking the square root on both sides 
\begin{equation}
e^{i\psi _{0}}=i\sigma e^{i\phi _{0}},  \label{psi(phi) 2}
\end{equation}
where $\sigma =\pm 1$. Equivalently 
\begin{equation}
\psi _{0}=\phi _{0}+\sigma \frac{\pi }{2}, \quad  {\rm mod}(2\pi ).
\label{psi0(phi0)}
\end{equation}
The most general form of $S_{0}$ is thus 
\begin{equation}
S_{0}=-\left[ 
\begin{array}{cr}
\sqrt{1-\tau _{0}}\, & i\sigma \sqrt{\tau _{0}}\, \\ 
i\sigma \sqrt{\tau _{0}}\, & \sqrt{1-\tau _{0}}\,
\end{array}
\right] e^{2i\phi _{0}},  \label{S0 polar 1}
\end{equation}
written in terms of the independent parameters $\tau _{0}$, $\phi _{0}$ and
the discrete variable $\sigma $, which have the range of variation 
\begin{mathletters}
\begin{eqnarray}
\tau _{0} & \in & [0,1], \\
\phi _{0} & \in & [0,2\pi ], \\
\sigma  & = & \pm 1.
\end{eqnarray} \label{range 0}
\end{mathletters}

From (\ref{rotation})-(\ref{s+-}), the matrix $S_{0}$
can be diagonalized by a $\pi /4$ rotation to give 
\begin{equation}
S_{0}^{\prime }=\left[ 
\begin{array}{cr}
e^{i\theta _{0}^{(+)}} & 0 \\ 
0 & e^{i\theta _{0}^{(-)}}
\end{array}
\right] \,,  \label{S0prime 1}
\end{equation}
where 
\begin{equation}
e^{i\theta _{0}^{(\pm )}}=r_{0}\pm t_{0}=-e^{2i\phi _{0}\pm i\sigma \beta
_{0}}  \label{theta+-}
\end{equation}
and 
\begin{equation}
\beta _{0}=\tan ^{-1}\sqrt{\frac{\tau _{0}}{1-\tau _{0}}},\qquad -\frac{\pi 
}{2}\leq \beta _{0}\leq \frac{\pi }{2}.  \label{beta}
\end{equation}
With the range of variation (\ref{range 0}) for $\tau _{0}$, $\phi _{0}$ and 
$\sigma $, $e^{i\theta _{0}^{(+)}}$ and $e^{i\theta _{0}^{(-)}}$ cover {\it %
twice} the torus defined by the two angles $\theta _{0}^{(+)}$,
$\theta_{0}^{(-)}$.

Eq. (\ref{theta+-}) is a transformation from the parameters $\tau _{0}$, $%
\phi _{0}$ and $\sigma $ to the parameters $\theta _{0}^{(+)}$, $\theta
_{0}^{(-)}$, whose Jacobian can be written as
\begin{equation}
\frac{1}{2}\frac{d\theta _{0}^{(+)}}{2\pi }\frac{d\theta _{0}^{(-)}}{2\pi }=%
\frac{1}{2}\frac{\,d\tau _{0}}{\pi \sqrt{\tau _{0}(1-\tau _{0})}}\,\frac{%
d\phi _{0}}{2\pi }.  \label{jacobian}
\end{equation}
Both sides of this last equation integrate to $1$ if the left-hand side is
integrated in the region $\theta _{0}^{(+)},\theta _{0}^{(-)}\in [0,2\pi ]$
and multiplied by $2$ to account for the fact that the region is visited
twice, and the right-hand side is integrated in the region specified by (\ref
{range 0}).

According to Eq. (\ref{invmeasLR}), the left-hand side of Eq. (\ref{jacobian}%
) represents the invariant measure for $S_{0}$ matrices with LR symmetry. A
function $f(\tau _{0},\phi _{0},\sigma )$ can be translated into a function $%
\widetilde{f}(\theta _{0}^{(+)},\theta _{0}^{(-)})$ using the transformation
(\ref{theta+-}); its average over the $S_{0}$ invariant measure can thus be written
as
\begin{eqnarray}
&&\int_{0}^{2\pi }\frac{d\theta _{0}^{(+)}}{2\pi }\int_{0}^{2\pi }\frac{%
d\theta _{0}^{(-)}}{2\pi }\widetilde{f}(\theta _{0}^{(+)},\theta _{0}^{(-)})
\nonumber \\
&=&\frac{1}{2}\sum_{\sigma =\pm 1}\int_{0}^{1}\frac{\,d\tau _{0}}{\pi \sqrt{%
\tau _{0}(1-\tau _{0})}}\int_{0}^{2\pi }\,\frac{d\phi _{0}}{2\pi }f(\tau
_{0},\phi _{0},\sigma ).  \label{av(f)}
\end{eqnarray}   
Here, on the left-hand side we integrate over the torus $\theta_0^{(+)}$,
$\theta_0^{(-)}$ only {\it once}. Suppose now that we are given
a function $F(\tau_0,\phi_0,\psi_0)=$ $F'(\tau_0,\phi_0,e^{i\psi_0})$ of the
{\it three parameters} appearing in Eq. (\ref{S0 polar}) and we want to compute its average over the above measure. First, we
make use of Eq. (\ref{psi(phi) 2}) to eliminate $\psi _{0}$ and write 
\begin{eqnarray}
F(\tau_0,\phi_0,\psi_0) & = &
F'(\tau_0,\phi_0,e^{i\psi_0}) =
F'(\tau_0,\phi_0,i\sigma e^{i\phi_0}) \nonumber \\ & = &
f(\tau_0,\phi_0,\sigma ) =
\widetilde{f}(\theta_0^{(+)},\theta_0^{(-)}),
\label{Fs}
\end{eqnarray}
where $f(\tau_0,\phi_0,\sigma )$ and
$\widetilde{f}(\theta_0^{(+)},\theta_0^{(-)})$, have the same meaning as in
(\ref{av(f)}) above. The average of this function can thus be written as in
(\ref{av(f)}) and subsequently as
\begin{eqnarray}
& & \frac{1}{2}\sum_{\sigma =\pm 1}\int_0^1 \frac{d\tau_0}
{\pi \sqrt{\tau_0 (1-\tau_0)}} \int_0^{2\pi} \,
\frac{d\phi_0}{2\pi }F'(\tau_0,\phi_0,i\sigma e^{i\phi_0}) \nonumber \\
& = & \frac{1}{2}\int_{0}^{1}\frac{d\tau _{0}}{\pi \sqrt{\tau _{0}
(1-\tau _{0})}} \int_{0}^{2\pi }\,\frac{d\phi _{0}}{2\pi }\int_{0}^{2\pi } 
d\psi _{0} \nonumber \\
& \times & \left[ \delta \left( \psi _{0}-\phi _{0}-\frac{\pi }{2} \right) +
\delta \left( \psi _{0}-\phi _{0}-3\frac{\pi }{2} \right) \right] \, 
F(\tau _{0},\phi _{0},\psi _{0}) , \nonumber \\ & &
\label{av(F) 1}
\end{eqnarray}
where we have used Eq. (\ref{psi0(phi0)}). Comparing the left hand-side of
(\ref{av(f)}) to the right-hand side of (\ref{av(F) 1}) we thus write
\begin{eqnarray}
\frac{ d{\theta^{(+)}_0} }{ 2\pi } \, \frac{ d\theta^{(-)}_0 }{ 2\pi } 
& \sim & \frac{2 \left[ \delta \left( \psi _{0}-\phi _{0}-\frac{\pi }{2} 
\right) + 
\delta \left( \psi _{0}-\phi _{0}-3\frac{ \pi }{2} \right) \right] }  
{\sqrt{1-\tau _{0}}} \nonumber \\ & \times & 
\frac{d\tau _{0}}{2\sqrt{\tau _{0}}} \, 
\frac{d\phi _{0}}{2\pi } \, \frac{d\psi _{0}}{2\pi } ,   
\label{dmuhat(dmu) 1}
\end{eqnarray} 
where the symbol $\sim $ indicates that the two measures are equivalent when
the left and right-hand sides are used to integrate the functions $%
\widetilde{f}(\theta _{0}^{(+)},\theta _{0}^{(-)})$ and
$F(\tau _{0},\phi _{0},\psi _{0})$, respectively, defined above. Obviously,
the angles in the argument of the delta functions above are defined modulo $%
2\pi $. As we have already noticed, the left-hand side of Eq.  
(\ref{dmuhat(dmu) 1}) is the invariant measure $d\widehat{\mu }(S_{0})$  
for scattering matrices $S_{0}$ of the form (\ref{S0prime 1}), i.e. for a  
LR-symmetric cavity. On the other hand, Eq. (\ref{inv meas N1}) shows that  
the last line of Eq. (\ref{dmuhat(dmu) 1}) is the invariant
measure $d\mu (S_{0})$ for scattering matrices $S_{0}$ of the more general  
form (\ref{S0 polar}). The relation between the two measures is thus
\begin{equation}
d\widehat{\mu }(S_{0}) \sim 
\frac{2 \left[ \delta \left( \psi _{0}-\phi _{0}-\frac{\pi }{2} \right) + 
\delta \left( \psi _{0}-\phi _{0}-3\frac{\pi }{2} \right) \right]}
{ \sqrt{1-\tau _{0}}} d\mu (S_{0}).  \label{dmuhat(dmu) 2}
\end{equation}
Here, the delta functions restrict the space of unitary and symmetric
matrices to the subspace of matrices of the form (\ref{S0 polar 1}).  

As was explained at the beginning of this section, we now express $d\mu
(S_{0})$ in terms of $d\mu (S)$ using Eq. (\ref{jacobian(S0,S)}). That
equation reads, for the present case,  
\begin{equation}
d\mu (S_{0})=\frac{\left[ \det \left( I_{2}-r_{b}r_{b}^{\dagger }\right)
\right] ^{3/2}}{\left| \det \left( I_{2}-Sr_{b}^{\dagger }\right) \right|
^{3}}\,d\mu (S).  \label{lrmeasure1}
\end{equation}
We substitute this last equation into Eq. (\ref{dmuhat(dmu) 2}) and use Eq. 
(\ref{inv meas N1}) to express $\,d\mu (S)$ in the polar representation. We
also note that the measure$\,\,d\widehat{\mu }(S_{0})$ appearing on the
left-hand side of Eq. (\ref{dmuhat(dmu) 2}), i.e. the differential
probability associated with the matrices $S_{0}$ [having the form  
(\ref{S0 N1})] for the LR-symmetric cavity, must coincide with the 
differential
probability $d\widehat{P}_{r_{b}}(S)$ we are looking for, associated with
the transformed matrices $S$ [having the form (\ref{S polar N1}), but with
the appropriate restrictions], i.e.
\begin{equation}
d\widehat{P}_{r_{b}}(S)=d{\hat{\mu}}(S_{0}).  \label{dhatP(S)dmu(S0)}
\end{equation} 

We thus have 
\begin{eqnarray}
d \widehat{P}_{r_b}(S) \, & \sim & 2 \, 
\frac{ \delta \left( \psi_0-\phi_0 - \frac{ \pi }{2} \right) + 
\delta \left( \psi_0 - \phi_0 - 3\frac{ \pi }{2} \right) }{ \sqrt{1-\tau_0} }
\nonumber \\
& \times & \frac{\left[ \det \left( I_{2}-r_{b}r_{b}^{\dagger }\right) 
\right]^{3/2}}{\left| \det \left( I_{2}-Sr_{b}^{\dagger }\right) 
\right| ^{3}} \,
\frac{d\tau }{2\sqrt{\tau }}\,\frac{d\phi }{2\pi } \, \frac{d\psi }{2\pi }.
\label{dpsrb1}
\end{eqnarray}
There remains to express the variables $\psi _{0}$, $\phi _{0}$, $\tau _{0}$
appearing in the delta-function arguments in terms of $\psi $, $\phi $, $
\tau$. This is done in App. \ref{tauphipsi(tauphipsi0)} for the particular 
case in which barrier 1 is transparent, so that its scattering matrix 
$S_{1}$ of Eq. (\ref{S1}) is the Pauli matrix $\sigma _{x}$, and barrier 2 
is described by Eq. (\ref{S2}) with {\it real} matrix elements. The result 
is 
\begin{equation} 
d\widehat{P}_{0,r_{2}^{\prime }}(S)\sim p_{r_{2}^{\prime }}(\tau ,\phi ,\psi
)\,d\tau \,d\phi \,d\psi ,
\label{lrptauphipsi 0}
\end{equation}
with   
\begin{eqnarray}
&& p_{r_{2}^{\prime }}(\tau ,\phi ,\psi ) = 
\frac{\left( 1-{r_{2}^{\prime }}^{2}\right)^{3/2}\left| 
\sqrt{1-\tau }-r_{2}^{\prime }e^{2i\phi }\right| }{ (2\pi)^{2}\sqrt{\tau } 
\left| \sqrt{1-\tau }\left( 1-{r_{2}^{\prime } }^{2}\right) -r_{2}^{\prime }
\tau e^{2i\phi }\right| ^{2}}  \nonumber \\
&& \times \left[ \delta \left( \psi -\phi -\alpha (\phi )-\frac{\pi }{2}
\right) +\delta \left( \psi -\phi -\alpha (\phi )-3\frac{\pi }{2}\right)
\right] , \nonumber \\ && \label{lrptauphipsi}
\end{eqnarray}
$\alpha(\phi) $ being given by Eq. (\ref{alpha}). We recall that the angles in
the arguments of the delta functions are defined modulo $2\pi$.  

\begin{figure}[tbp]
\input epsf \epsfxsize=8.5cm \centerline{\epsffile{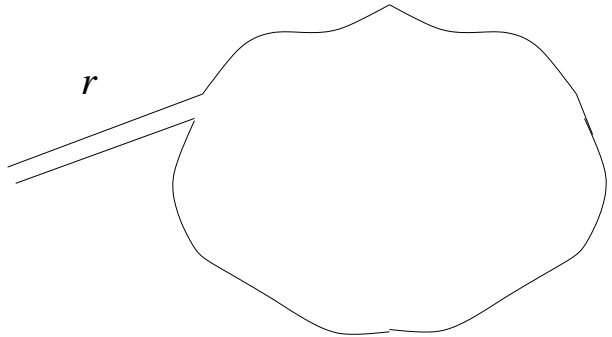}}
\caption{A ballistic chaotic cavity with reflection symmetry connected to just
one lead, supporting one open channel, in the absence of direct processes.
The 1-dimensional $S$ matrix $r=-e^{2i\phi}$ is distributed according to the
invariant measure.}
\label{cavitylrbroken}
\end{figure}

As a first check, set $r_{2}^{\prime }=0$, corresponding to the case of no
barriers. We obtain
\begin{eqnarray}
&& d\widehat{P}_{0,0}(S) \sim \frac{d\tau }{\pi \sqrt{\tau 
\left( 1-\tau \right) }} \, \frac{d\phi }{2\pi } \, \nonumber \\ & \times & 
\frac{1}{2}\left[ \delta \left( \psi -\phi -\frac{\pi }{2}\right)
+\delta \left( \psi -\phi -3\frac{\pi }{2}\right) \right] d\psi .
\label{dPhat00}
\end{eqnarray}
Thanks to the delta functions, we recover the situation of LR symmetry. As
expected, the right-hand side of Eq. (\ref{dPhat00}) is the invariant
measure defined for that symmetry, Eq. (\ref{dmuhat(dmu) 1}). As a second
check, we analyze the case $r_{2}^{\prime }\rightarrow -1$, that corresponds
to obstructing the waveguide on the right. We show in Appendix
\ref{ProofLRdP(0,-1)(S)} that (\ref{lrptauphipsi 0}) gives in this case
\begin{equation}
d\widehat{P}_{0,-1}\left( S \right) \sim \delta \left( \tau \right) d\tau
\frac{d\phi }{2\pi } \frac{1}{2}\left[ \delta \left( \psi -\frac{\pi }{2}
\right) +\delta \left( \psi - 3\frac{\pi }{2}\right) \right] d\psi .
\label{LRdP(0,-1)(S)}
\end{equation}
The conductance distribution reduces to a one-sided delta function at zero, 
as it should. Notice that the variable $\phi $ is uniformly distributed in  
the two extreme cases $r'_2=0$ and $r'_2=-1$; this is not so for an 
arbitrary value of $r_{2}^{\prime }$. In the limiting case $r'_2=-1$ we end 
up with a LR-symmetric ballistic cavity connected to just one lead
(see Fig. \ref{cavitylrbroken}): the resulting $1$-dimensional $S$ matrix, i.e.
$r=-e^{2i\phi }$, is distributed according to the {\it invariant measure}:
there is thus {\it no memory left of the LR symmetry of the cavity}. In fact,
the right-hand side of Eq. (\ref{LRdP(0,-1)(S)}) is identical to that of
Sect. \ref{AScavities}, Eq. (\ref{ASdP(0,-1)(S)}), for an AS cavity with the
right waveguide obstructed. As we shall see later on, in
Sec. \ref{asymmetric position}, this is a peculiarity of the 1-channel case.

\begin{figure}[ht]
\input epsf \epsfxsize=8.5cm \centerline{\epsffile{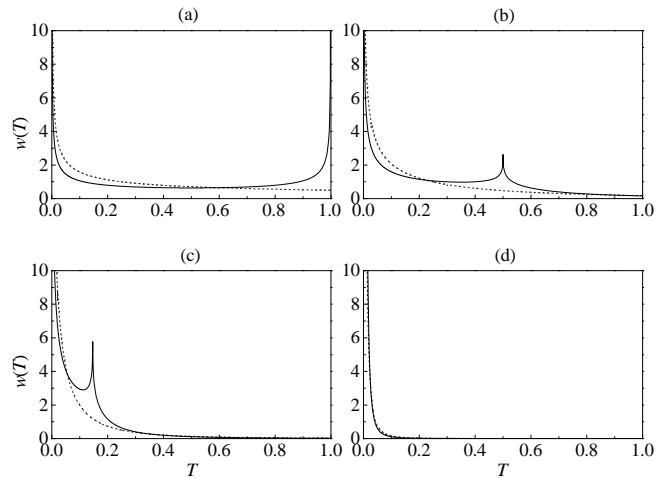}}
\caption{The heavy line shows the evolution of the $T$ distribution obtained
by numerical integration over $\phi$ of Eq. (\ref{q(tauphi)r2}) for a LR
symmetric cavity and one barrier defined by $r'_2 = -\cos \epsilon$, as a
function of the parameter $\epsilon$. Panels (a), (b), (c), (d) show the cases
$\epsilon=\frac{\pi}{2}, \frac{\pi}{4}, \frac{\pi}{8}, \frac{\pi}{32}$,
respectively. The dotted lines show, for comparison, the distributions that
correspond to an AS cavity and the same barrier as for the corresponding
heavy lines.}
\label{wtsim0rp2}
\end{figure}

To get the joint distribution of $\tau $ and $\phi $ for arbitrary  
$r_{2}^{\prime }$ we integrate (\ref{lrptauphipsi}) over $\psi $. We find 
\begin{eqnarray}
q_{r_{2}^{\prime }}(\tau ,\phi ) & = & \frac{1}{2\pi ^{2}} \,  
\frac{\left( 1-{r_{2}^{\prime }}^{2}\right) ^{3/2}}{\sqrt{\tau }} 
\nonumber \\ & \times &  
\frac{\left| \sqrt{1-\tau}-r_{2}^{\prime } \, e^{2i\phi }\right| }{\left| 
\sqrt{1-\tau }\left( 1-{r_{2}^{\prime }}^{2}\right) -r_{2}^{\prime }\tau \, 
e^{2i\phi }\right| ^{2}}.
\label{q(tauphi)r2}
\end{eqnarray} 
The $T=\tau$ distribution $w(T)$ is obtained by integrating
$q_{r'_2}(\tau, \phi)$ over $\phi$. Fig. \ref{wtsim0rp2} shows (heavy lines)
the evolution of $w(T)$ with the parameter $r'_2$. In the same figure we
compare that sequence of distributions with those corresponding to an AS  
cavity (dotted lines) with the same $r'_2$. In the former case {\it the
system ``remembers'' in a rather conspicuous way that, although the resulting
configuration is asymmetric, the cavity has LR symmetry}.

\section{Breaking the reflection symmetry of cavities with an asymmetric
position of the waveguides}
\label{asymmetric position} 

\begin{figure} 
\input epsf \epsfxsize=8.5cm \centerline{\epsffile{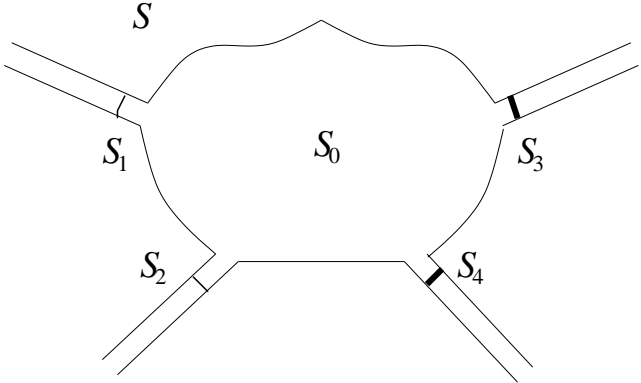}} 
\caption{A ballistic cavity connected to four waveguides symmetrically 
located, by means of four, in principle different, barriers.} 
\label{four asymm} 
\end{figure}

In the present section we study the effect of external mixing of LR symmetry 
in the absence of direct processes, i.e. for $\langle S\rangle=0$: the problem 
will be that of a LR-symmetric cavity connected to two asymmetrically 
positioned waveguides in the absence of barriers. We proceed as follows. We 
first consider the LR-symmetric cavity connected to four symmetrically 
positioned waveguides (each supporting one open channel) by means of four, in 
general different, barriers, as shown in Fig. \ref{four asymm}. The two  
barriers on the left side are then removed, while those on the right are made 
perfect reflectors. 

We call $S_0$ the matrix associated with the LR-symmetric cavity connected to 
the four symmetrically located waveguides in the {\it absence} of barriers. 
The matrix $S$ in the presence of the four barriers is then given by Eq. 
(\ref{S0(S)}), i.e.  
\begin{equation}
S=r_b + t'_b \,\frac{1}{ I_4 - S_0 \, r'_b } \, S_0 \, t_b \, , \label{S(S0)} 
\end{equation}
where 
\begin{equation} 
t_b = t'_b = \left( \begin{array}{cccc} 
t_1 & 0 & 0 & 0 \\  0 & t_2 & 0 & 0 \\ 0 & 0 & t_3 & 0 \\ 0 & 0 & 0 & t_4 
\end{array} \right) 
\end{equation} 
\begin{equation} 
r_b = \left( \begin{array}{cccc} 
r_1 & 0 & 0 & 0 \\ 0 & r_2 & 0 & 0 \\ 0 & 0 & r'_3 & 0 \\ 0 & 0 & 0 & r'_4 
\end{array} \right) \, , \qquad 
r'_b = \left( \begin{array}{cccc} 
r'_1 & 0 & 0 & 0 \\ 0 & r'_2 & 0 & 0 \\ 0 & 0 & r_3 & 0 \\ 0 & 0 & 0 & r_4 
\end{array} \right) \, . 
\end{equation}  

As explained above, we now open the left waveguides and block the right ones  
by means of perfect reflectors, so that  
\begin{equation} 
t_b =t'_b= \left( 
\begin{array}{cr} I_2 & 0_2 \\ 0_2 & 0_2 \end{array} 
\right) \, , \qquad 
r_b =r'_b= \left( 
\begin{array}{cr} 0_2 & 0_2 \\ 0_2 & -I_2 \end{array} 
\right) \, , 
\end{equation} 
where $I_2$ and $0_2$ denote the two-dimensional unit and zero matrices,
respectively. 

The $4\times 4$ matrix $S_0$ has the structure (\ref{lrs}), i.e.
\begin{equation} 
S_0 = \left( \begin{array}{cr} r_0 & t_0 \\ t_0 & r_0 \end{array} \right) \, , 
\end{equation} 
where $r_0$ and $t_0$ are 2-dimensional matrices. The matrix $S$ of Eq.  
(\ref{S(S0)}) then reads  
\begin{equation} 
S = \left( \begin{array}{cr} 
r_0 - t_0 \, \frac{1}{ I_2 + r_0 } \, t_0 & 0 \\ 0 & -I_2 
\end{array} \right) \, . 
\end{equation} 

\begin{figure}
\input epsf \epsfxsize=8.5cm \centerline{\epsffile{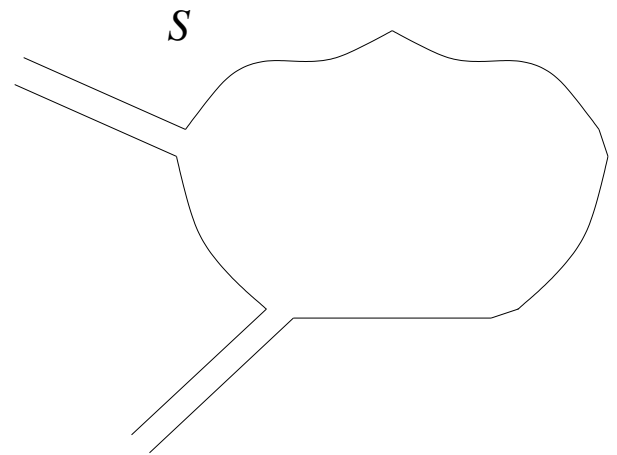}}
\caption{A ballistic chaotic cavity connected to two asymmetrically
located waveguides, without direct processes. The 2-dimensional $S$ matrix is
{\it not} distributed according to the invariant measure, but it is close.}
\label{two asymm}
\end{figure}

The 1-1 block of the above expression is the $2\times 2$ scattering matrix 
of the final system consisting of a LR-symmetric ballistic cavity connected  
to two waveguides on the left (see Fig. \ref{two asymm}), i.e
\begin{equation} 
s = r_0 - t_0 \, \frac{1}{ I_2 + r_0 } \, t_0 \, . \label{s(r0,t0)} 
\end{equation}

Using the result (\ref{s+-}) we can express $r_0$ and $t_0$ as 
\begin{equation} 
r_0 = \frac{1}{2} \left[ s^{(+)} + s^{(-)} \right] \, , \label{r0(s+-)} 
\end{equation} 
\begin{equation} 
t_0 = \frac{1}{2} \left[ s^{(+)} - s^{(-)} \right] \, , \label{t0(s+-)} 
\end{equation} 
where $s^{(\pm)}$ are $2\times 2$ unitary and symmetric matrices. 

A numerical calculation was performed, in which 4-dimensional $S_0$ matrices 
were generated with a distribution corresponding to their invariant measure: 
this was done by constructing an ensemble of $s^{(\pm)}$ matrices, Eqs. 
(\ref{r0(s+-)}), (\ref{t0(s+-)}), distributed as two independent $COE$'s. 
From Eq. (\ref{s(r0,t0)}), the resulting $S$ matrices were then evaluated. 

\begin{figure}
\input epsf \epsfxsize=8cm \centerline{\epsffile{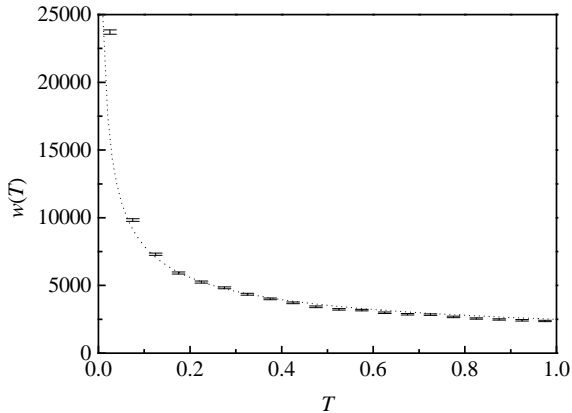}}
\caption{The $T$-distribution for a LR symmetric cavity connected to two
asymmetrically located waveguides. The dotted line correspond to an AS cavity
connected to two waveguides.}
\label{extmix}
\end{figure}

The distribution of the resulting transmission coefficient $T$ is shown in 
Fig. \ref{extmix}. For comparison, the distribution $1/2\sqrt{T}$  
corresponding to an AS cavity connected to two one-channel waveguide and with 
$\langle S\rangle =0$ is shown with dotted line. Although the LR-symmetric cavity with
external symmetry breaking has a $T$ distribution very close to $1/2\sqrt{T}$, 
there is a statistically significant deviation which indicates that the 
resulting system has a memory of the point symmetry of the cavity. This is to 
be contrasted with the result mentioned right before Eq. (\ref{q(tauphi)r2}) 
for the single 1-channel cavity illustrated in Fig. \ref{cavitylrbroken}.

\section{Results and conclusions}

One of the main purposes of the present paper has been the extension of 
previous studies on transport through ballistic chaotic cavities with reflection
symmetry to include the presence of direct processes. In Sect. \ref{LR
directprocesses} we treated the problem of {\it fully left-right (LR) symmetric
systems} in the presence of direct processes. The statistical distribution of
the $S$ matrix, found analytically in Eq. (\ref{lrkernel2}), consists of the
product of two Poisson's kernels with the optical matrices
$\langle s^{(+)}\rangle$ and $\langle s^{(-)}\rangle$, respectively. For no
direct transmission processes, $\langle t\rangle =0$, and real direct 
reflections $\langle r\rangle$, we calculated analytically the distribution 
of the transmission coefficient $w(T)$ for the one-channel case. The 
difference with the $T$ distribution for an asymmetric cavity (AS) with the same
optical matrix $\langle S\rangle$, which is striking for 
$\langle r\rangle=0$, becomes less dramatic as $|\langle r\rangle|$ 
increases: that evolution is shown in Fig. \ref{wtsimrr}.
 
The other main purpose of this work has been the study of LR-symmetry 
breaking by an asymmetric coupling of a LR-symmetric cavity to the outside.
Two ways of producing ``external mixing" of the spatial symmetry were 
analyzed:

\paragraph{} \label{one}
 In Sect. \ref{snolr} we studied the effect of {\it breaking the
reflection symmetry of a cavity by direct processes}. The system consists of a
ballistic cavity with reflection symmetry connected to two symmetrically
positioned waveguides by means of barriers which, in general, are allowed to be
different (Fig. \ref{cavitylr}). We found analytically, in Eqs.
(\ref{lrptauphipsi 0}) and (\ref{lrptauphipsi}), the statistical distribution of
the $S$ matrix for the one-channel case in each waveguide and, for simplicity,
when only the barrier in the right waveguide is
present ($r'_2 \neq 0$). The $T$ distribution is strikingly different from that for the fully AS case (i.e 
the one in which the cavity itself is AS) having the same optical 
$\langle S\rangle$ matrix, as shown in Fig. \ref{wtsim0rp2} for various 
values of $r'_2 \neq 0$. We conclude that this two-waveguide system,
although asymmetric with respect to the LR operation, {\it has a memory of 
the reflection symmetry of the
cavity} from which it is constructed. In the
limit $r'_2 \longrightarrow -1$ the right waveguide is blocked and we end up 
with a LR-symmetric ballistic cavity connected, without any barrier, to
just one lead, supporting one open channel (see Fig. \ref{cavitylrbroken}).
We found that the resulting
1-dimensional matrix $S=e^{i\theta}$ is distributed
according to its invariant measure (i.e., $\theta $ is uniformly distributed)
and, as a result, has {\it no memory left of the LR symmetry of the cavity}:
this was found, though, to be a peculiarity of the one-waveguide--one-channel
case (in fact, see the end of next paragraph).

\paragraph{} \label{two}
In Sect. \ref{asymmetric position} we studied, {\it in the absence of
direct processes, the effect of
external mixing of LR symmetry induced by an asymmetric position of
the waveguides}. The result is a LR-symmetric cavity
connected, without any barriers, to two waveguides on its left-hand side
(see Fig. \ref{two asymm}).
Let $T$ denote the total transmission coefficient between those two 
waveguides; its distribution $w(T)$ was calculated numerically for the 
one-channel case in each waveguide and compared, in Fig. \ref{extmix}, with 
$1/2\sqrt{T}$, the $T$ distribution arising from the invariant measure 
$d\mu ^{(\beta =1)}(S)$ for AS systems. Although the difference between the
two distributions is quite small, it is statistically significant. This 
problem is clearly equivalent to having, on one side of the cavity, just one 
waveguide (coupled to the cavity without any barrier) supporting two open 
channels. In this one-waveguide--two-channel problem the resulting $S$ 
matrix is thus distributed very closely to its invariant measure, the 
difference exhibiting {\it some memory left of
the reflection symmetry of the cavity}.

Two additional points are worth mentioning. First, from an experimental point of
view, we notice that microwave cavities and acoustic systems might represent
good possibilities to study the interplay between the symmetry of the cavity and
external mixing in the statistical distribution of the conductance of such a
structure. Finally, the problem described in \ref{two} above is relevant to the
study of transport between two one-channel leads connected by a ``double" Cayley
tree \cite{moisesthesis}. In fact, under suitable circumstances the two problems
can be mapped unto each other. This problem will be reported elsewhere.

\acknowledgments

One of the authors (M. M.) wishes to acknowledge supports by DGAPA-UNAM,
and CONACyT, M\'exico.

\appendix 

\section{Derivation of Eq. (2.29)}
\label{ProofASdP(0,-1)(S)}

We saw in Sec. \ref{AScavities} that the distribution
$dP_{\langle S\rangle}(S)$ of the
scattering matrix of a cavity connected to two waveguides, where the one on the right
of the cavity has a barrier, is given by 
\begin{equation}
dP_{0,r'_2}(S) = \frac{ \left( 1 - |r'_2|^2 \right)^{3/2} } 
{ \left| 1 - \sqrt{1-\tau} \, e^{2i\psi} \, {r'_2}^* \right|^3 } \, 
\frac{d\tau}{2\sqrt{\tau}} \, \frac{d\phi}{2\pi} \, \frac{d\psi}{2\pi}.
\end{equation}

To see the behaviour of $dP_{0,r'_2}$ for $r'_2=-1$, let $r'_2$ be a real 
number: assume for simplicity $r^{\prime}_2=-\cos\epsilon$; we are 
interested in the limit $\epsilon\longrightarrow 0$. Also, let us introduce the
positive parameter $\eta\ll 1$ in order to avoid the integrable singularity 
in $\tau$. Of
course, we will take the limit $\eta\longrightarrow 0$ later on. Because the variable $\phi$ is uniformly distributed, the joint 
probability distribution of $\tau$ and $\psi$ can be written as 
\begin{equation} 
p_{\eta,\epsilon}(\tau, \psi) = \frac{C_{\eta}}{4\pi \sqrt{\tau+\eta^2}} 
\frac{\left| \sin\epsilon \right|^3 } {\left| 1 + \cos\epsilon\sqrt{1-\tau} 
\, e^{2i\psi} \right|^3 }
, \label{p(eta,epsilon)(tau,psi)}
\end{equation}
where $C_{\eta}$ is a normalization constant which depends on the parameter 
$\eta$. 

We have the following properties of $p_{\eta,\epsilon}(\tau,\psi)$: 

\begin{enumerate} 

\item From (\ref{p(eta,epsilon)(tau,psi)})
we see that \begin{equation}
p_{0,0}(\tau,\psi) = \lim_{\eta\rightarrow 0} \,
\lim_{\epsilon\rightarrow 0} p_{\eta,\epsilon}(\tau,\psi) = 0
\end{equation}
for all $\tau$ and $\psi$, except for $\tau = 0$ and 
$\psi = \frac{\pi}{2},3\frac{\pi}{2}$, where the denominator is zero 
\begin{equation}
\left| 1 + \sqrt{1-\tau}\, e^{2i\psi} \right|^3 = 0
.\end{equation}

\item For $\tau=0$ and $\psi=\frac{\pi}{2},3\frac{\pi}{2}$ we have 
\begin{eqnarray}
&& p_{0,0} \left( \tau=0,\psi=\frac{\pi}{2},3\frac{\pi}{2} \right) 
\nonumber \\ && = \lim_{\eta\rightarrow 0} \, \lim_{\epsilon\rightarrow 0} 
\, p_{\eta,\epsilon} \left( \tau=0, \psi = \frac{\pi}{2},3\frac{\pi}{2} 
\right) \nonumber \\ && = \lim _{\eta \rightarrow 0} \, 
\lim_{\epsilon \rightarrow 0} \, \frac{C_{\eta}}{4\pi\eta} \, \left| \cot 
\frac{\epsilon}{2} \right|^3 \longrightarrow \infty
\end{eqnarray}

\item The function $p_{0,0}(\tau,\psi)$ is normalized to the 
unity: 
\begin{eqnarray}
&& \int_0^1 d\tau \int_0^{2\pi} d\psi \,\, p_{0,0}(\tau, \psi) 
\nonumber \\ && = \lim_{\eta,\epsilon\rightarrow 0}  \int_0^1 d\tau 
\int_0^{2\pi} d\psi \,\, p_{\eta,\epsilon}(\tau, \psi) = 1
\end{eqnarray}

\end{enumerate}

Then the only function which satisfies those conditions is 
\begin{equation}
p_{0,0}(\tau, \psi) = \delta \left(\tau\right) \, \frac{1}{2} \left[ \delta
\left( \psi-\frac{\pi}{2} \right) + \delta \left( \psi-\frac{3\pi}{2}
\right) \right] .
\end{equation}

Finally, the distribution of the $S$ matrix in the above limit is given by
Eq. (\ref{ASdP(0,-1)(S)}).

\section{Derivation of Eq. (3.11)}
\label{DerivEqw(T)LRav(r) 1} 

For $\langle r\rangle$ {\it real} and $s^{(\pm)} = e^{i\theta_{\pm}}$,
Eq. (\ref{PkernLRt0N1}) can be written as
\begin{equation}
d\widehat{P}_{\langle r\rangle }(S) = 
\frac{ 1 - \left\langle
r \right\rangle^2 } { \left|
1 - \left\langle r\right\rangle e^{i\theta_{+}} \right| ^{2} } \, \frac{ 1 - \left\langle
r \right\rangle^2 } { \left|
1 - \left\langle r\right\rangle e^{i\theta_{-}} \right| ^{2} } \, \frac{d\theta_{+}}{2\pi} \, \frac{d\theta_{-}}{2\pi} . 
\label{dhatP(r)(S)fullLRN1} 
\end{equation} 

The transmission amplitude is given by [see Eq. (\ref{s+-})]: 
\begin{equation} 
t = \frac{1}{2}\left( e^{i\theta_+} - e^{i\theta_-} \right),
\end{equation} 
and the transmission coefficient is written as 
\begin{equation} 
T = \left| t \right|^2 = 
\frac{1}{2}\left[ 1 - \cos \left( \theta_+ - \theta_- \right) \right].
\end{equation} 
The $T$ distribution $w_{\langle r\rangle}(T)$ is obtained from 
\begin{equation} 
w_{\langle r\rangle}(T) = \int \delta\left(T - \frac{1}{2}\left[ 1 - \cos 
\left( \theta_+ - \theta_- \right) \right] \right) 
d\widehat{P}_{\langle r\rangle}(S). 
\label{w(r)(T)fullLRN1} 
\end{equation} 

We make the following change of variables in order to solve the integral:
\begin{equation} 
\begin{array}{cc} 
\theta & = \frac{1}{2} \left( \theta_+ - \theta_- \right), \\ 
\theta'& = \frac{1}{2} \left( \theta_+ - \theta_- \right); 
\end{array}
\end{equation}
the range of variation are: for $\theta'\in (0, 2\pi)$, 
$\theta\in (-\theta', \theta')$ and for $\theta'\in (\pi, 2\pi)$, 
$\theta\in (-2\pi+\theta', 2\pi-\theta')$.

Substituting (\ref{dhatP(r)(S)fullLRN1}) in (\ref{w(r)(T)fullLRN1}),
considering the fact that the integrand is an even function of $\theta$ and
writing the delta function in terms of its roots in the variable $\theta$ we have
\begin{equation} 
\delta\left(T-\sin^2\theta\right) = \frac{1}{2\sqrt{T\left(1-T\right)}} 
\left[ \delta\left( \theta - \theta_1 \right) + 
\delta\left( \theta - \theta_2 \right) \right] ,
\end{equation}
where $\theta_2 = \pi - \theta_1$ and $\theta_1 = \arcsin \sqrt{T}$; 
finally, after some algebra, $w_{\langle r\rangle}(T)$ can be written as a 
sum of two terms: 
\begin{equation} 
w_{\langle r\rangle}(T) = 
\frac{ \left( 1 - \left\langle r\right\rangle^2 \right)^2 } 
{ \pi^2 \sqrt{T\left(1-T\right)} } 
\left[ I_1 \left( T, \left\langle r\right\rangle \right) + 
I_2 \left( T, \left\langle r\right\rangle \right) \right],
\label{w(T)I1I2}
\end{equation} 
where, for $k = 1, 2$, 
\begin{eqnarray} 
&& I_k \left( T, \left\langle r\right\rangle \right) \nonumber \\ 
& = & \int_0^{\pi} d\theta' \int_0^{\theta'} d\theta 
\frac{1}{ \left[ \left(1+\left\langle r\right\rangle^2\right) - 
2\left\langle r\right\rangle \cos \left(\theta' + \theta\right) \right] } 
\nonumber \\ &\times & 
\frac{ \delta\left( \theta - \theta_k \right) } 
{ \left[ \left(1+\left\langle r\right\rangle^2\right) - 
2\left\langle r\right\rangle \cos \left(\theta' - \theta\right) \right] }. 
\end{eqnarray}

Again, after some algebra the sum of the two integrals give a single one:
\begin{equation} 
I_1 \left( T, \left\langle r\right\rangle \right) + 
I_2 \left( T, \left\langle r\right\rangle \right) = \frac{1}{c} 
\int_0^{\pi} \frac{ d\theta' }{ a - b\cos \theta' + \cos^2 \theta' } , 
\end{equation}
where 
\begin{eqnarray} 
a & = & \frac{1}{c} \left[ \left( 1 + \left\langle r\right\rangle^2 
\right)^2 - 4\left\langle r\right\rangle^2 T \right], \nonumber \\ 
b & = & \frac{4}{c} \left\langle r\right\rangle 
\left( 1+\left\langle r\right\rangle^2 \right) \sqrt{1-T}, 
\label{smallabc} \\ 
c & = & 4\left\langle r \right\rangle^2. \nonumber 
\end{eqnarray}
Now, making the change of variable $x=\cos\theta'$, (\ref{w(T)I1I2}) can be
written as
\begin{equation} 
w_{\langle r\rangle}(T) = 
\frac{ \left( 1 - \left\langle r\right\rangle^2 \right)^2 } 
{ 4\left\langle r\right\rangle^2 \pi^2 \sqrt{T\left(1-T\right)} } 
\left[ I_+ \left( T, \left\langle r\right\rangle \right) + 
I_- \left( T, \left\langle r\right\rangle \right) \right], 
\label{w(r)(T)I+I-}
\end{equation}
where now
\begin{equation}
I_{\pm}\left(T,\left\langle r\right\rangle\right) = 
\int_0^1 \frac{dx}{ \sqrt{1-x^2}\left(a \pm bx + x^2 \right) }. \label{I+-}
\end{equation} 

By means of a change of variables 
\begin{mathletters}
\begin{equation} 
u  = - \frac{x + \left(A+B\right)}{x + \left(A-B\right)}, 
\end{equation} 
\begin{equation}
v  = - \frac{x - \left(A+B\right)}{x - \left(A-B\right)},
\end{equation}
\end{mathletters} 
where 
\begin{eqnarray}
A & = & \frac{1}{b}\left(1+a\right), \\ 
B & = &\frac{1}{b}\sqrt{\left(1+a\right)^2-b^2}, \label{capsB}
\end{eqnarray}
the indefinite integrals, $Indef_{\pm}$, corresponding to each one of the  
above, can be transformed to 
\begin{mathletters}
\begin{equation}
Indef_+ = \frac{2B}{\sqrt{C}D} \int \frac{ \left| u + 1 \right| } 
{ \sqrt{u^2+p}\left( u^2+q \right) }du, \label{xu}
\end{equation}
\begin{equation}
Indef_- = - \frac{2B}{\sqrt{C}D} \int \frac{ \left| v + 1 \right| } 
{ \sqrt{v^2+p}\left( v^2+q \right) }dv. \label{xv}
\end{equation}
\end{mathletters}
where 
\begin{eqnarray}
p & = & \frac{a-b\left(B+A\right)+\left(B+A\right)^2} 
{ a+b\left(B-A\right)+\left(B-A\right)^2 } \label{smallp} \\
q & = & \frac{1-\left(B+A\right)^2}{1-\left(B-A\right)^2} \label{smallq}
\end{eqnarray}
and 
\begin{equation}
\begin{array}{cl}
C & = 1 - (B-A)^2 , \\ 
D & =  a + b(B-A) + (B-A)^2. 
\end{array}\label{capsCD}
\end{equation}

Although the integrals (\ref{I+-}) seem to give the same result under the 
change $b\rightarrow -b$, they do not, because the cutoff $x_u=B-A$ in
(\ref{xu}), and $x_v=A-B$ in (\ref{xv}), are different. One must be careful 
evaluating the integrals in the limits. The results are 
\begin{mathletters} 
\begin{eqnarray}
I_+ \left( T,\left\langle r\right\rangle \right) &=& 
\frac{ 2B }{ \sqrt{C}D\sqrt{p-q} }
\Biggl[ \arctan \left( \frac{ 1-\left\langle r\right\rangle^2 }
{ 2\left\langle r\right\rangle \sqrt{1-T} } \right) \nonumber \\ &-& 
\frac{ 1 }{ 2\sqrt{p} } \ln \left( 
\frac{ 1+\left\langle r\right\rangle^2 + 
2\left\langle r\right\rangle \sqrt{T} }
{ 1+\left\langle r\right\rangle^2-2\left\langle r\right\rangle \sqrt{T} }
\right) \Biggr],  
\end{eqnarray}
\begin{eqnarray}
I_- \left( T,\left\langle r\right\rangle \right) &=& 
\frac{ 2B }{ \sqrt{C}D\sqrt{p-q} }
\Biggl[ \pi -\arctan \left( \frac{ 1-\left\langle r\right\rangle^2 }
{ 2\left\langle r\right\rangle \sqrt{1-T} } \right) \nonumber \\ &+&  
\frac{ 1 }{ 2\sqrt{p} } \ln \left( 
\frac{ 1+\left\langle r\right\rangle^2 + 
2\left\langle r\right\rangle \sqrt{T} }
{ 1+\left\langle r\right\rangle^2-2\left\langle r\right\rangle \sqrt{T} }
\right) \Biggr].  
\end{eqnarray}
\label{I+I-}
\end{mathletters}

Now, we substitute the sum of equations (\ref{I+I-}) in (\ref{w(r)(T)I+I-}) 
to obtain the result
\begin{equation} 
w_{\langle r\rangle}(T) = 
\frac{ \left( 1 - \left\langle r\right\rangle^2 \right)^2 } 
{ 4\left\langle r\right\rangle^2 \pi^2 \sqrt{T\left(1-T\right)} } 
\frac{ 2\pi B }{ \sqrt{C} D \sqrt{p-q} }; 
\end{equation}
using Eqs. (\ref{smallabc}), (\ref{capsB}), (\ref{smallp}), 
(\ref{smallq}) and (\ref{capsCD}) the final result (\ref{w(T)LRav(r) 1}) is
obtained.

\section{Derivation of Eqs. (4.23), (4.24)}
\label{tauphipsi(tauphipsi0)}

For the particular case in which barrier 1 is transparent [see Fig. 
\ref{cavitylr}], so that its scattering matrix $S_{1}$ of Eq. (\ref{S1}) is 
the Pauli matrix $\sigma _{x}$, and barrier 2 is described by Eq. 
(\ref{S2}) with {\it real} matrix elements, Eq. (\ref{dpsrb1}) can be written
as
\begin{eqnarray}
d\widehat{P}_{0,r'_2}(S) \, & \sim & 2 \,
\frac{ \delta \left( \psi_0-\phi_0 - \frac{ \pi }{2} \right) +  
\delta \left( \psi_0 - \phi_0 - 3\frac{ \pi }{2} \right) }{\sqrt{1-\tau_0}}
\nonumber \\ 
& \times & \frac{ \left( 1-{r'_2}^2 \right)^{3/2} }
{ \left| 1- \sqrt{1-\tau} \, r'_2 \, e^{2i\psi} \right|^3 } \, 
\frac{d\tau}{2\sqrt{\tau }}\,\frac{d\phi }{2\pi } \,  \frac{d\psi }{2\pi }. 
\label{dpsrb2}
\end{eqnarray}
Also, the transformation $S_0(S)$ given by Eq. (\ref{lrs02}) can be written in 
terms of its elements as follows: 
\begin{eqnarray}
r_0 & = & \frac{ 1 }{ 1-r'_2 \, r'} \, 
\left[ r \, ( 1-r'_2
\, r') + r'_2 \, t^2 \right], \nonumber \\ r'_0 & = & \frac{ 1 }{ 1-r'_2 \, r'} \, \left(
r'- r'_2 \right), \label{r0r0pt02} \\ 
t_0 & = & \frac{ 1 }{ 1-r'_2 \, r'} \, t_2 \, t
, \nonumber \end{eqnarray}
or in terms of the independent parameters [see Eqs.(\ref{S polar N1}) and 
(\ref{S0 polar})] as:
\begin{mathletters}
\begin{eqnarray}
\sqrt{1-\tau _0}\,e^{2i\phi _0}&=&e^{2i\phi }\,\frac{\sqrt{1-\tau }-r'_2\,
e^{2i\psi }}{1-r'_2\sqrt{1-\tau }\,e^{2i\psi} } \label{r0} \\
\sqrt{1-\tau _0}\,e^{2i\psi _0}&=&e^{2i\psi }\,\frac{\sqrt{1-\tau }-r'_2\,
e^{-2i\psi }}{1-r'_2\sqrt{1-\tau }\,e^{2i\psi }}
\label{r0p} \\\sqrt{\tau _0}\,e^{i(\phi _0+\psi _0)}&=&\frac{t_2\sqrt{\tau }\,
e^{i(\phi +\psi )}}{1-r'_2\sqrt{1-\tau }\,e^{2i\psi }} \label{t0}
\end{eqnarray}
\end{mathletters}

From (\ref{r0}) or (\ref{r0p}) we find 
\begin{equation}
\sqrt{1-\tau^{(0)}}=\frac{\left| \sqrt{1-\tau } - r'_2 \, e^{2i\psi} \right|}
{\left| 1-\sqrt{1-\tau}\, r'_2 \, e^{2i\psi} \right| }; \label{tau0}
\end{equation}
also, dividing the (\ref{r0}) by (\ref{r0p}) we obtain
\begin{equation}
e^{2i(\psi _0 -\phi _0)}=e^{2i(\psi -\phi )}\,\frac{\sqrt{1-\tau }-r'_2\,
e^{-2i\psi }}{\sqrt{1-\tau }-r'_2\,e^{2i\psi }}.
\label{psi0mphi0}
\end{equation}
Because the roots of the delta functions appearing in Eq. (\ref{dpsrb1}) 
satisfy $e^{2i(\psi
_0-\phi _0)}=-1$, from (\ref{psi0mphi0}) we find\begin{equation}
e^{2i\psi }=-e^{2i\phi }\,e^{2i\alpha (\phi )}  \label{psi}
\end{equation}
where 
\begin{equation}
e^{i\alpha (\phi )}=\frac{\sqrt{1-\tau }-r'_2\,e^{-2i\phi }}
{\left| \sqrt{1-\tau }-r'_2\,e^{2i\phi }\right| }. \label{alpha}
\end{equation}
Then, we have the conditions for $\psi $: 
\begin{equation}
\begin{array}{ll}
\psi -\phi -\alpha (\phi )=\frac{\pi }{2} \quad & {\rm for} \quad 
\psi
_{0}-\phi _{0}=\frac{\pi }{2} \\ 
\psi -\phi -\alpha (\phi )=3\frac{\pi }{2} \quad & {\rm for}\quad 
\psi
_{0}-\phi _{0}=3\frac{\pi }{2}.
\end{array}
\end{equation}

The Jacobian for the transformation $\psi_0\longrightarrow \psi$ is
\begin{equation}
\left| \frac{\partial }{\partial \psi }\left( \psi _{0}-\phi _{0}\right)
\right| =\frac{\left| (1-\tau )-{r_{2}^{\prime }}^{2}\right| }{\left| \sqrt{%
1-\tau }-r_{2}^{\prime }\,e^{-2i\psi }\right| ^{2}} .  
\end{equation}
Then we write 
\begin{eqnarray}
& & \delta \left( \psi _0-\phi _0-\frac{2n+1}{2}\pi 
\right) \nonumber \\ &=&\frac{\left| \sqrt{1-\tau }-r_{2}^{\prime }\,e^{-2i\psi
 }\right| ^{2}}{\left|(1-\tau )-{r_{2}^{\prime }}^{2}\right| } 
\delta \left[ \psi -\phi -\alpha (\phi )-\frac{2n+1}{2}\pi \right] ,
\nonumber \\ & &
\label{deltapsi0-phi0} 
\end{eqnarray}
for $n=0,1$. 

From (\ref{psi}) and (\ref{alpha}) we find 
\begin{eqnarray}
\sqrt{1-\tau} - r'_2 \, e^{2i\psi } & = & 
\frac{(1-\tau )-{r'_2}
^{2}}{\sqrt{1-\tau} - r'_2 \,e^{2i\phi}}
 \,, \label{taupsi(phi) 1} \\ 
1-r'_2 \sqrt{ 1-\tau } e^{2i\psi } & = & 
\frac{ \sqrt{ 1-\tau } \left(1-{r'_2}^2 \right) - r'_2 \tau \, e^{2i\phi} }
{ \sqrt{ 1-\tau } \, r'_2 \, e^{2i\phi} }
. \label{taupsi(phi) 2}\end{eqnarray}
Finally, substituting Eqs. (\ref{tau0}), (\ref{deltapsi0-phi0}),
(\ref{taupsi(phi) 1}) and (\ref{taupsi(phi) 2}) in Eq. (\ref{dpsrb2}), we
arrive to 
\begin{equation}
d\widehat{P}_{0,r'_2}(\tau, \phi, \psi) \sim p_{r'_2}(\tau ,\phi ,\psi) \, 
d\tau
\, d\phi \, d\psi ,\end{equation}
where $p_{r'_2}(\tau ,\phi ,\psi)$ is given by Eq. (\ref{lrptauphipsi}). 

\section{Derivation of Eq. (4.26)}
\label{ProofLRdP(0,-1)(S)}

In Sect. \ref{snolr} we find the joint distribution of $\tau$, $\phi$ and
$\psi$ [Eq. (\ref{lrptauphipsi})]. From that it is easy to integrate over
$\psi$ to find the joint distribution of $\tau$ and $\phi$ to be
\begin{equation}
q_{r'_2 }(\tau ,\phi) =
\frac{\left( 1-{r'_2 }^{2}\right)^{3/2}\left|
\sqrt{1-\tau }-r'_2 e^{2i\phi }\right| }{ (2\pi)^{2}\sqrt{\tau }
\left| \sqrt{1-\tau }\left( 1-{r'_2  }^{2}\right) -r'_2
\tau e^{2i\phi }\right| ^{2}}
\end{equation}

As in App. \ref{ProofASdP(0,-1)(S)} we assume for simplicity
$r'_2=-\cos\epsilon$; again we introduce the parameter $\eta\ll 1$. Of
course, we will take the limits $\eta$, $\epsilon \longrightarrow 0$; then
\begin{equation}
q_{\eta,\epsilon}(\tau, \phi) = \frac{C_{\eta}}{2\pi^2}
\frac{|\sin\epsilon|^3}{\sqrt{\tau + \eta^2}}
\frac{\left| \sqrt{ 1-\tau }+ \cos\epsilon \, e^{2i\phi} \right| }
{\left| \sqrt{ 1-\tau } \sin^2\epsilon + \tau \cos\epsilon \,
e^{2i\phi} \right|^2 }. \label{lrptauphiapp2}
\end{equation}
where $C_{\eta}$ is a normalization constant which depends on $\eta$.

Again, as before we have the following properties for
$q_{\eta, \epsilon}(\tau, \phi)$:

\begin{enumerate}

\item From (\ref{lrptauphiapp2}) we see that
\begin{equation}
q_{0,0}(\tau, \phi) = \lim_{\eta\rightarrow 0} \,
\lim_{\epsilon\rightarrow 0} \, q_{\eta,\epsilon}(\tau,\phi) = 0
\end{equation}
for all $\tau$ and $\phi$, except for $\tau=0$ and
$\phi=\frac{\pi}{2},3\frac{\pi}{2}$, where the denominatror is zero:
\begin{equation}
\left| \sqrt{ 1-\tau } \sin^2\epsilon + \tau \cos\epsilon \,
e^{2i\phi} \right|^2 = 0
\end{equation}

\item $\tau\neq 0$ and $\forall \phi$.

It is easy to see from (\ref{lrptauphiapp2}) that in this case 
\begin{equation}
q_{0,0}(\tau\neq 0, \phi) = \lim_{\eta} \, \lim_{\epsilon\rightarrow 0} \,
q_{\eta,\epsilon}(\tau,\phi) = 0.
\end{equation}

\item For $\tau = 0$ and $\phi = \frac{\pi}{2}, 3\frac{\pi}{2}$ we have
\begin{eqnarray}
&& q_{0,0} \left( \tau=0, \phi=\frac{\pi}{2},3\frac{\pi}{2} \right)
\nonumber \\
&&= \lim_{\eta\rightarrow 0}\, \lim_{\epsilon\rightarrow 0} \,
q_{\eta,\epsilon} \left(\tau=0,\phi=\frac{\pi}{2},3\frac{\pi}{2}\right)
\nonumber \\
&&= \lim_{\eta\rightarrow 0}\, \lim_{\epsilon\rightarrow 0} \,
\frac{C_{\eta}}{2\pi^2\eta} \, \left| \tan \frac{\epsilon}{2}
\right|
\end{eqnarray}

\item  For $\tau=0$, $\phi\neq \frac{\pi}{2}, 3\frac{\pi}{2}$ we obtain
\begin{eqnarray}
&& q_{0,0} \left( \tau=0, \phi\neq\frac{\pi}{2},3\frac{\pi}{2} \right)
\nonumber \\
&&=\lim_{\eta\rightarrow 0} \, \lim_{\epsilon\rightarrow 0} \,
q_{\eta,\epsilon} \left(\tau=0, \phi\neq\frac{\pi}{2},3\frac{\pi}{2}
\right) \nonumber \\ && = \lim_{\eta\rightarrow 0} \,
\lim_{\epsilon\rightarrow 0} \, \frac{C_{\eta }}{2\pi ^{2}\eta}\,
\frac{\left| 1+\cos \epsilon \,e^{2i\phi}\right| }
{\left| \sin\epsilon \right| } \longrightarrow \infty .
\end{eqnarray}

\item Also, the function $q_{0,0}(\tau,\phi)$ is normalized to the unity:
\begin{eqnarray}
&&\int_0^1 d\tau \int_0^{2\pi} q_{0,0}(\tau,\phi) \nonumber \\
&&=\lim_{\eta,\epsilon \rightarrow 0} \int_0^1 d\tau \int_0^{2\pi}
q_{\eta,\epsilon}(\tau,\phi) \, d\tau \, d\phi=1 .
\end{eqnarray}

These conditions defines the function
\begin{equation}
q_{0,0} \left(\tau, \phi \right) = \delta \left( \tau \right) \,
\frac{1}{2\pi } .
\end{equation}
We thus arrive at Eq. (\ref{LRdP(0,-1)(S)}).

\end{enumerate}

\end{multicols}

\end{document}